%% file: O2DCPaper.tex
\documentclass[10pt]{iopart}

\usepackage[lmargin=0.7in,rmargin=0.7in,tmargin=.7in,bmargin=.7in]{geometry}
\usepackage{iopams}
\usepackage{graphicx} % for pdf, bitmapped graphics files
\usepackage[font=footnotesize,labelfont=bf]{caption}
\usepackage{subcaption} % for subcaptions in side-by-side figures
\usepackage{cite}
\usepackage{hyperref} % for linking the citations
\usepackage{xcolor}
\usepackage{booktabs}
\usepackage{longtable}
\usepackage{float}
\usepackage{changepage}

\makeatletter
\def\endthebibliography{%
  \def\@noitemerr{\@latex@warning{Empty `thebibliography' environment}}%
  \endlist
}
\makeatother

\begin{document}

\title[A reaction mechanism for oxygen plasmas]{A reaction mechanism for oxygen plasmas}

\author{Tiago C Dias$^1$, Chloé Fromentin$^1$, Luís L Alves$^1$, Antonio Tejero-del-Caz$^{1,2}$, Tiago Silva$^1$ and Vasco Guerra$^1$}

\address{$^1$Instituto de Plasmas e Fusão Nuclear, Instituto Superior Técnico, Universidade de Lisboa, Portugal\\
		 $^2$Departamento de Física, Facultad de Ciencias, Universidad de Córdoba, Campus de Rabanales, Spain}
\ead{tiago.cunha.dias@tecnico.ulisboa.pt}
\vspace{10pt}
\begin{indented}
\item[] April 2023
\end{indented}

\begin{abstract}
This work presents a reaction mechanism for oxygen plasmas, i.e. a set of reactions and corresponding rate coefficients that are validated against benchmark experiments. The kinetic scheme is validated in a DC glow discharge for gas pressures of 0.2--10 Torr and currents of 10--40 mA, using the 0D LisbOn KInetics (LoKI) simulation tool and available experimental data. The comparison comprises not only the densities of the main species in the discharge - $\mathrm{O_2(X^3\Sigma_g^-)}$, $\mathrm{O_2(a^1\Delta_g)}$, $\mathrm{O_2(b^1\Sigma_g^+)}$ and $\mathrm{O(^3P)}$ - but also the self-consistent calculation of the reduced electric field and the gas temperature. The main processes involved in the creation and destruction of these species are identified. 
Moreover, the results show that the oxygen atoms play a dominant role in gas heating, via recombination at the wall and quenching of $\mathrm{O_2(X^3\Sigma_g^-,v)}$ vibrations and $\mathrm{O_2}$ electronically-excited states. It is argued that the development and validation of kinetic schemes for plasma chemistry should adopt a paradigm based on the comparison against \textit{standard validation tests}, as it is done in electron swarm validation of cross sections.
\end{abstract} 

\vspace{1pc}
 \vspace{1pc}\\
\submitto{\PSST}
%\maketitle

%\pagebreak
%\tableofcontents

\vspace{-1cm}
\section{Introduction}\label{sec:introduction}

Low-temperature oxygen-containing plasmas have been used in a manifold of applications. They give an important contribution in various fields such as semiconductor manufacturing~\cite{Cvelbar_2005,Childres_2011}, thin film synthesis~\cite{Park_2013} and plasma medicine~\cite{Bauer_2016,Sousa_2011,Schroter_2020}. The kinetics of reactive oxygen species have been studied extensively also due to their relevance, e.g., in oxygen-iodine lasers~\cite{Braginskiy_2005,Guerra_2010}, plasma-assisted combustion~\cite{Ju_2015}, CO$_2$ conversion~\cite{Morillo-Candas_2019,Chen_2020,Guerra_2022}, NO production~\cite{Douat_2016} and atmospheric chemistry studies~\cite{Azyazov_2015}. 
The main strength of oxygen plasmas is the inherent ability of generating large quantities of highly reactive species, such as $\mathrm{O(^3P)}$, $\mathrm{O_2(a^1\Delta_g)}$, $\mathrm{O_2(b^1\Sigma_g^+)}$ and O$_3$, with a relatively small energy input. The high reactivity comes together with great complexity and, although these plasmas were vastly studied, important open questions remain pending.

The understanding of O$_2$ discharges requires a combined effort of modelling and experimental work. On the one hand, experiments allow to access and characterize real plasma systems. However,  very often the amount of physical parameters that can be accessed by experimental diagnostics is limited and the interpretation of the experimental measurements is not straightforward. On the other hand, modelling provides information on both the microscopic and macroscopic scales and access to physical quantities not directly available from measurements.

One of the pillars of low-temperature plasma (LTP) modelling is the kinetic scheme describing the interactions between the heavy species and their connection with the electrons. This defines not only the densities of heavy species but also the power required to sustain the discharge, via the electron energy distribution function (EEDF), which is strongly correlated with the composition of the gas mixture. Therefore, the definition of a \textit{reaction mechanism}, i.e., a set of reactions and corresponding rate coefficients that are validated against
benchmark experiments, is mandatory to make sensible modelling predictions of the system under study.  By definition, a benchmark is a reference point against which things can be compared. In the present context, benchmark experiments correspond to a significant ensemble of experimental data, intended (or suited) for model validation, obtained in well defined and reproducible conditions, using established diagnostics, and assessing multiple quantities.

The development of reaction mechanisms is a concept introduced by the combustion community, GRI-Mech constituting a well-known example~\cite{Grimech}. However, this procedure is still somewhat uncommon in the LTP community, where extensive validation of electron-impact cross sections by swarm analysis is well established~\cite{Pancheshnyi_2012,Pitchford_2013,Alves_2016,Pinhao_2020} but the validation of plasma-chemistry models against benchmark experiments remains to some extent elusive. In fact, although many modellers follow the procedure of comparing their results against experiment for many years, a community consensus on recommended practices and definition of experiments as golden standards to assess the results of the models is still missing~\cite{Alves_2023_foundations}. Following~\cite{Alves_2023_foundations}, such a definition requires a combined effort of experimentalists and modellers and would enhance our ability to understand the rationale of the approximations made and bring systematic consistency between models. Regarding $\mathrm{O_2}$ discharges, there are various kinetic schemes available in the literature \cite{Annusova_2018,Dias_2020,Pietanza_2020,Murakami_2013,Toneli_2015,Lieberman_1995,Ionin_2007,Braginskiy_2005}. Nonetheless, to the best of our knowledge, none of them can be considered to be validated against a comprehensive set of dedicated experiments designed as a benchmark.

In recent years, a group from the Laboratoire de Physique des Plasmas (LPP) in Paris, in collaboration with a team from the Lomonosov Moscow State University (MSU), performed a systematic and thorough experimental work concerning the kinetics of $\mathrm{O(^3P)}$~\cite{Booth_2019}, $\mathrm{O_2(a^1\Delta_g)}$~\cite{Booth_2020} and $\mathrm{O_2(b^1\Sigma_g^+)}$~\cite{Booth_2022}. The studies were done in a DC glow discharge in pure O$_2$ for gas pressures of 0.2--10 Torr and currents of 10--40 mA. The experimental campaigns revealed new mechanisms for the formation and quenching of $\mathrm{O_2(a^1\Delta_g)}$ and $\mathrm{O_2(b^1\Sigma_g^+)}$, pointed out the elementary processes underlying heterogeneous O-atom recombination, and made available a very complete set of experimental data to be used for the validation of kinetic schemes.

Our group in Lisbon developed for several years a kinetic scheme for O$_2$ kinetics that describes rather well different sets of experiments \cite{Guerra_1997,Guerra_1999,Kutasi_2010,Marinov_2013,Annusova_2018}. The latest published work \cite{Annusova_2018} consists on the study of inductively coupled plasma discharges, where the influence of the vibrational kinetics of $\mathrm{O_2(X^3\Sigma_g^-,v)}$ was also included. However, a preliminary comparison with the experimental measurements made at LPP and MSU has shown that our model was still failing to capture important trends, specially for the kinetics of $\mathrm{O_2(a^1\Delta_g)}$ and $\mathrm{O_2(b^1\Sigma_g^+)}$, which motivated the present investigation.

In this work, we leverage on recent experimental data~\cite{Booth_2019,Booth_2020,Booth_2022} from LPP and MSU to develop a reaction mechanism for oxygen plasmas. The validation is made using a self-consistent 0-D model describing the positive column of a DC glow discharge. This discharge is well suited for unraveling the kinetics of oxygen species, as it generates in many conditions an axially homogeneous plasma and is accessible to various diagnostics. The densities of the main species - $\mathrm{O(^3P)}$,  $\mathrm{O_2(X^3\Sigma_g^-)}$, $\mathrm{O_2(a^1\Delta_g)}$ and $\mathrm{O_2(b^1\Sigma_g^+)}$ - are benchmarked against experiment and the main processes of species production/destruction are enlightened. Moreover, the validation is extended to the prediction of the gas temperature and the reduced maintenance electric field (the ratio of the electric field to the gas density). Bear in mind that in some configurations DC glow discharges may exhibit striations in the axial direction~\cite{Swain_1971,Tsendin_2010,Boeuf_2022,Kolobov_2022}, depending on the type of gas, pressure and tube dimensions, which turns their accurate modelling deeply challenging, but in the experimental works of \cite{Booth_2019,Booth_2020,Booth_2022} no striations are mentioned.

The calculations in this work are performed with the LisbOn KInetics (LoKI) 0D simulation tool, constituted by two modules: a Boltzmann solver (LoKI-B) for the two-term electron Boltzmann equation \cite{Tejero-del-Caz_2019,Tejero-del-Caz_2021} and a chemical solver (LoKI-C) for the heavy-species kinetics \cite{Guerra_2019,NPRiME_website}. In past works, LoKI has been used to describe with success the positive column of DC glow discharges in different gas mixtures~\cite{Guerra_2019,Silva_2020,Fromentin_2023_CO2O2,Fromentin_2023_CO2N2}. However, other authors have employed different models to describe the kinetics of glow discharges. For instance, glow discharges have been studied with fluid models, where the spatial dependence of the electron kinetics is accounted for by means of the local-energy or local-field approximations~\cite{Booth_2019,Booth_2022,Ingold_1997,Petrov_1999,Franklin_2000,Alves_2007,Laca_2019}. The spatial variation of the electron energy distribution can be considered also with the so-called ``nonlocal'' approach~\cite{Bernstein_1954,Tsendin_1995,Kolobov_1995,Kortshagen_1996,Pfau_1996,Uhrlandt_1999,Zobnin_2009}, in conditions where the energy relaxation length is much higher than the tube radius, which is not the case for the conditions examined in this work~\cite{Ivanov_2000,Bogdanov_2003}. A more correct, complex and time-consuming formulation describes the spatial variation of the electron kinetics by solving the space-dependent electron Boltzmann equation~\cite{Bush_1995,Alves_1997,Arndt_2001,Uhrlandt_2002,Loffhagen_2009,Grubert_2013,Yuan_2017}. Here, we use the simpler 0D approach, which was verified to provide a good description of the system under study for most species (O$_3$ being an exception) by comparison with 1D calculations \cite{Viegas_2023}, and is further confirmed in section \ref{sec:validation}. Another important point is the use of the two-term assumption for a Legendre expansion of the electron distribution function in the velocity space~\cite{Phelps_1985_twoTermApprox}, since it may not be accurate when the inelastic collisions are strong or the electric field is high~\cite{White_2003}. Nevertheless, for the  reduced electric fields typical in these discharges, $\lesssim 100$~Td, the two-term approximation should provide good results for the case of oxygen~\cite{Dias_2023}.

The paper is organized as follows. Section~\ref{sec:methods} presents (i) the discharge set-up under study, (ii) the self-consistent 0D kinetic model, (iii) the reactions comprising the kinetic scheme, focusing on the differences with previous works, and (iv) the additional input data required to simulate oxygen plasmas. Section \ref{sec:results} presents the validation of the model by comparison with experiment and analyzes the main processes controlling the discharge behavior. Finally, section \ref{sec:conclusions} summarizes the main findings of this paper and sets out future work.

\section{Methods and conditions}\label{sec:methods}

\subsection{Glow discharge set-up}\label{sec:dischargeSetup}
The experimental set-up of the glow discharge simulated in this work is described in detail in \cite{Booth_2019,Booth_2020,Booth_2022}. The DC discharge is ignited in a Pyrex tube 56 cm long with an internal radius of 1 cm. The electrodes are located in side-arms separated by 52.5 cm, so as to fill the main tube only with the positive column. The surface temperature of the tube is kept constant at 50$^o$C = 323.15~K using a thermostatic bath. The temperature difference between the inner and outer surfaces of the tube is less than 2 K and therefore negliglible \cite{Booth_2019}. The gas pressure varies from 0.2 to 10 Torr, the discharge current from 10 to 40 mA and the flow rate is pressure dependent, being 2 sccm at 0.2 Torr and 10 sccm at 10 Torr. The oxygen recombination probabilities at the wall were measured in \cite{Booth_2020} and are used as input to the model, cf. section \ref{sec:kineticScheme}.

\subsection{0D global model LoKI}\label{sec:LoKI}
The calculations in this work were performed with the LoKI 0D simulation tool, constituted by two modules: a Boltzmann solver (LoKI-B) for the two-term electron Boltzmann equation \cite{Tejero-del-Caz_2019,Tejero-del-Caz_2021} and a chemical solver (LoKI-C) for the heavy-species kinetics \cite{Guerra_2019,NPRiME_website}. This section gives an overview of the two modules and how they are coupled in a self-consistent way.

\subsubsection{LoKI-B}\label{sec:LoKIB}\quad\\
LoKI-B \cite{Tejero-del-Caz_2019} solves a space-independent form of the electron Boltzmann equation under the usual two-term assumption, for a Legendre expansion of the electron distribution function in the velocity space \cite{Phelps_1985_twoTermApprox}. The simulation tool addresses the electron kinetics in any complex gas mixture, considering inelastic and superelastic collisions with any target state. On input, the code requires the operating conditions (reduced electric field, gas temperature and pressure), the gas mixture, the populations of the internal states and the relevant sets of electron-scattering cross sections. The cross sections can be obtained from the open-access website LXCat \cite{LXCat_website} or other sources, and should be organized in datafiles compliant with the LXCat format. On output, it provides (i) the EEDF, (ii) the electron power absorbed from the electric field and transferred to the different collisional channels, (iii) the electron swarm parameters, such as the mobility and the diffusion coefficient and (iv) the electron-impact rate-coefficients calculated from the integration of the corresponding cross sections over the EEDF. This information is then used in the chemical solver as follows. 

\subsubsection{LoKI-C}\label{sec:LoKIC}\quad\\
LoKI-C solves a system of 0D rate-balance equations for all the heavy-species densities~($n_s$) in the plasma, where the gain-loss terms due to chemical volume reactions ($S_{s}^\mathrm{chem}$), transport to the wall ($S_s^\mathrm{transp}$) and gas flow ($S_s^\mathrm{flow}$) are spatially averaged, according to
\begin{equation}\label{eq:rateBalance}
\frac{\partial n_s}{\partial t} = S_s^\mathrm{chem} + S_s^\mathrm{transp} + S_s^\mathrm{flow}\ .
\end{equation}
The term $S_{s}^\mathrm{chem}$ comprises electron-impact reactions, where the rate coefficients from LoKI-B are used, and chemical reactions between heavy particles (see section \ref{sec:kineticScheme}). 
The transport of neutral species to the wall is described with the model proposed by Chantry \cite{Chantry_1987}, as detailed in section 4.1 of Guerra~\textit{et~al.}~\cite{Guerra_2019}. This approach takes into account both the diffusion time of the species ($\tau_s^\mathrm{diff}$), from the reactor volume to the wall, and the probability ($\gamma_s$) of destruction at the wall.
The transport of positive ions is assumed to be given by ambipolar diffusion, as  described by the high plasma-density limit of the ion transport theories \cite{Phelps_1990,Coche_2016}. Since negative ions $\mathrm{O^-}$ are predominant in $\mathrm{O_2}$ discharges, their effect on the electron density profile is taken into account, following\cite{Guerra_1999}. The impact of different charged-particle transport theories on the discharge kinetics is discussed and quantified elsewhere~\cite{Alves_2023}. Note that the electron mobility and diffusion coefficient obtained in LoKI-B are used in the self-consistent calculation of the ambipolar diffusion coefficients. Finally, $S_s^\mathrm{flow}$ entails creation and destruction terms due to flow of particles entering and exiting the reactor. The inflow is defined by the experimental conditions and the outflow is determined so as to conserve the total number of atoms in the discharge volume (see~\cite{Silva_2020} for more details).

Take note that the rate-balance equation for the electrons is not solved since the calculations are performed for a given electron density consistent with the experimental discharge current, cf. section \ref{sec:LoKIBC}.

In addition to the equations for the species densities, LoKI-C also solves a thermal model for the self-consistent calculation of the gas temperature $T$ \cite{Pintassilgo_2014,Pintassilgo_2016}. Under constant pressure and assuming that heat conduction is the dominant cooling mechanism along the volume, the gas thermal balance equation writes
\begin{equation}\label{eq:heatTransferGeneral}
N C_p \frac{\partial T}{\partial t} = Q_\mathrm{in} - \lambda_\mathrm{g} \nabla^2T\ ,
\end{equation}
where $N$ is the gas density, $C_p$ is the heat capacity of the gas at constant pressure, $\lambda_\mathrm{g}$ is the thermal conductivity and $Q_\mathrm{in}$ is the total net power per unit volume transferred to gas heating. The latter term has two contributions: (i) the net power per unit volume transferred from  the electrons to the gas due to elastic collisions, obtained from LoKI-B, and (ii) the sum of all the reaction rates multiplied by the corresponding reaction enthalpies. For reactions occurring at the wall, it is assumed that half of the energy is accommodated by the wall, while the other half is released for gas heating. The impact of this assumption is extensively discussed by Pintassilgo~\textit{et~al.}~\cite{Pintassilgo_2014}. %$C_p$, $\lambda_\mathrm{g}$ and $Q_\mathrm{in}$ are assumed homogeneous and are calculated using the volume average values of the gas temperature and the species densities. The temperature-dependent values of $C_p$ and $\lambda_\mathrm{g}$ adopted in this work are presented in section~\ref{sec:transpThermoData}.

In conditions of axial homogeneity and stationarity, the solution of equation (\ref{eq:heatTransferGeneral}) corresponds to the following parabolic radial profile~\cite{Pintassilgo_2014}:
\begin{equation}\label{eq:Tradial}
T(r) \simeq T_0 - (T_0-T_\mathrm{nw}) \frac{r^2}{R^2}\ ,
\end{equation}
where $R$ is the tube radius, $r\in[0,R]$, and $T_0$ and $T_\mathrm{nw}$ are the gas temperatures at the center and near the wall, respectively. Assuming this parabolic profile, the temporal evolution of the average gas temperature $T_\mathrm{g} \equiv \langle T(r) \rangle_r = \frac{1}{2} (T_0 + T_\mathrm{nw})$ is described by the following equation:
\begin{equation}\label{eq:heatTransferAverage}
N C_p \frac{\partial T_\mathrm{g}}{\partial t} \simeq Q_\mathrm{in} - \frac{8 \lambda_\mathrm{g} (T_\mathrm{g}-T_\mathrm{nw})}{R^2}\ ,
\end{equation}
where $C_p$, $\lambda_\mathrm{g}$ and $Q_\mathrm{in}$ are assumed homogeneous and are calculated as a function of $T_\mathrm{g}$.
Equation (\ref{eq:heatTransferAverage}), for the gas temperature, is solved coupled to equations (\ref{eq:rateBalance}) for the species densities, with a value of $T_\mathrm{nw}$ to be determined. First, we assume that the temperature drop between the inner and outer walls of the reactor is negligible, in accordance with experiment (c.f. section \ref{sec:dischargeSetup}), such that a fixed inner-wall temperature $T_\mathrm{w}$ is taken, equal to the outer-wall temperature of 323.15 K. Second, we consider a convective heat flux between the gas and the inner-wall given by:
\begin{equation}\label{eq:fluxConv}
\Gamma_\mathrm{nw} = h_\mathrm{gas-wall} (T_\mathrm{nw}-T_\mathrm{w})\ ,
\end{equation}
where $h_\mathrm{gas-wall}$ is the convection coefficient between the gas and the wall. This heat flux matches the power per unit area that flows outward the gas by conduction, which can be estimated using the gas-temperature profile~(\ref{eq:Tradial}):
\begin{equation}\label{eq:fluxCond}
\Gamma_\mathrm{nw} = \lambda_\mathrm{g} |\nabla T| \simeq \frac{4 \lambda_\mathrm{g}}{R} (T_\mathrm{g} - T_\mathrm{nw})\ .
\end{equation}
Combining equations (\ref{eq:fluxConv}) and (\ref{eq:fluxCond}), we obtain:
\begin{equation}\label{eq:Tnw}
T_\mathrm{nw} = \frac{\frac{4\lambda_\mathrm{g}}{R} T_\mathrm{g} + h_\mathrm{gas-wall} T_\mathrm{w}}{\frac{4\lambda_\mathrm{g}}{R} + h_\mathrm{gas-wall}}
\end{equation}
The impact of different $h_\mathrm{gas-wall}$ values on the determination of $T_\mathrm{g}$ is studied in section \ref{sec:validation}. An alternative approach to obtain the temperature jump between $T_\mathrm{nw}$ and $T_\mathrm{w}$ would be to consider a thermal accommodation coefficient for the gas particles hitting the tube wall, that accounts for the energy accommodation by the wall in reactions occurring at the surface and in elastic collisions in a combined manner, as it is done e.g. in \cite{Abada_2002,Booth_2019}. Although a direct comparison between the coefficients used in either formulation is not straightforward, the similarity of the calculated temperature jumps shows they are compatible. Additionally, the parabolic profile of gas temperature considered here~\cite{Pintassilgo_2014,Pintassilgo_2016}, motivated by the solution of the thermal balance equation with constant thermal properties and heat source, could be replaced by a Bessel profile, motivated by the typical shape of the electron density profile~\cite{Laderman_1971,Tiedtke_1995,Yalin_2002}. It is verified \textit{a posteriori} that, under the conditions of the present study, the parabolic profile describes well the behavior evidenced in experiment (\textit{cf.} figure \ref{fig:TgasRadial30mA}). In any case, at higher pressures ($\gtrsim$ 100 Torr), not considered in this work, phenomena of discharge contraction may arise, where the temperature profile is neither parabolic nor Bessel, and a space-dependent model is required to provide an accurate description~\cite{Vialetto_2022,Golubovskii_2011}.

\subsubsection{Self-consistent coupling: LoKI-B+C}\label{sec:LoKIBC}
\begin{figure}[H]
	\centering
	\includegraphics[width=\columnwidth]{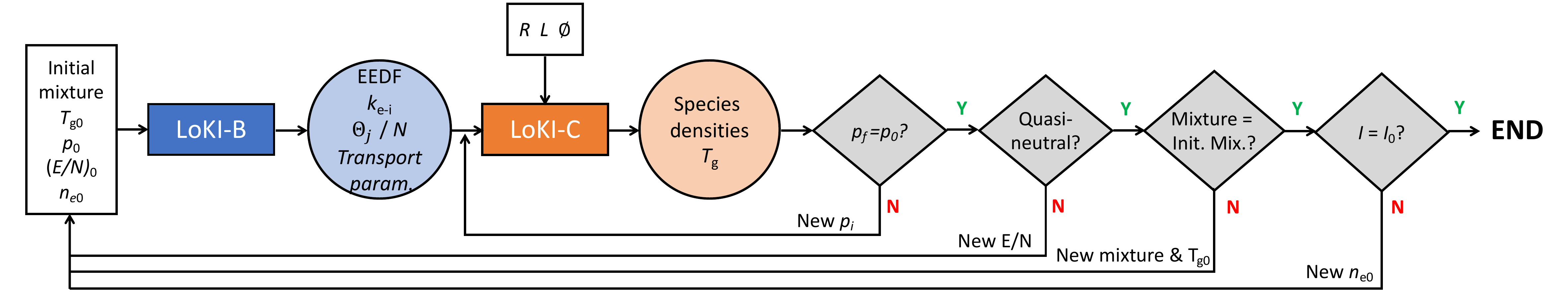}
	\caption{Workflow of LoKI, adapted from \cite{Silva_2020}. See text of section~\ref{sec:LoKIBC} for details on the notation used in the figure.}
	\label{fig:workflowLoKI}
\end{figure}

\quad\\
The workflow of LoKI is presented in various works \cite{Coche_2016,Guerra_2019,Silva_2020} and is schematized in figure~\ref{fig:workflowLoKI}. LoKI calculates a self-consistent solution of both electron and heavy-species kinetics in the plasma for user-defined working conditions: initial mixture composition ($[i]/N$), gas pressure ($p_0$), discharge current ($I_0$), reactor radius ($R$) and length ($L$), and flow rate ($\phi$). For initial guesses of the mixture $[i]/N$, average gas temperature ($T_\mathrm{g0}$), reduced electric field ($(E/N)_0$) and electron density ($n_{e0}$), the workflow starts with LoKI-B calculating the EEDF, the electron-impact rate coefficients ($k_\mathrm{e-i}$), the electron power transferred into the different collisional channels ($\Theta_j/N$) and the electron transport parameters. Then, this information is used in LoKI-C to calculate the species densities and the gas temperature until steady-state is reached. If the steady-state pressure ($p_f$) is not the same as the working pressure ($p_0$), the pressure at the beginning of the chemical calculation ($p_i = p(t=0)$) is varied until the condition is satisfied. Similarly, the $E/N$ used in the Boltzmann solver is adjusted so as to achieve quasi-neutrality in steady-state. Additionally, consistency in both modules is assured by updating the steady-state gas mixture and temperature calculated by LoKI-C into LoKI-B. Finally, the electron density is varied so as to obtain a calculated discharge current $I = e v_d \pi R^2 n_e$ equal to the experimental value given as input (here, $e$ is the electron charge and $v_d$ is the simulated drift velocity). Notice that, for other discharge configurations where the current cannot be given as input, the input power density may be used instead for the electron density calculation.
In this work, the relative tolerances for the pressure, quasi-neutrality, mixture composition and electron density cycles are $10^{-4}$, $10^{-3}$, $10^{-4}$ and $10^{-3}$, respectively.

\subsection{Reaction mechanism}\label{sec:kineticScheme}

The reaction mechanism presented here represents an update relatively to the work by Annu\v{s}ova~\textit{et~al.}~\cite{Annusova_2018}, and benefits from the recent experimental campaigns by Booth and co-workers \cite{Booth_2019,Booth_2020,Booth_2022} and insights from the MSU team~\cite{Vasiljeva_2004,Braginskiy_2005}. The current set of reactions considers the following charged and neutral species: e, $\mathrm{O_2^+}$,$\mathrm{O^+}$, $\mathrm{O^-}$, $\mathrm{O_2(X,v=0:41)}$, $\mathrm{O_2(a)}$, $\mathrm{O_2(b)}$, $\mathrm{O_2(Hz)}$, $\mathrm{O(^3P)}$, $\mathrm{O(^1D)}$, $\mathrm{O_3}$ and $\mathrm{O_3^*}$, where $\mathrm{O_2(Hz)}$ is an effective sum of the $\mathrm{O_2(A^{\prime 3}\Delta_u,A^3\Sigma_u^+,c^1\Sigma_u^-)}$ Herzberg states~\cite{Vasiljeva_2004} and $\mathrm{O_3^*}$ is an effective state of vibrationally-excited ozone~\cite{Marinov_2013}.

This section aims to summarize the reactions comprising the scheme, focusing on the differences relatively to the previous version~\cite{Annusova_2018}, namely: (i) quenching of $\mathrm{O_2(a)}$~\cite{Braginskiy_2005,Booth_2020} and $\mathrm{O_2(b)}$~\cite{Booth_2022}, (ii) inclusion of the $\mathrm{O_2(Hz)}$ species~\cite{Vasiljeva_2004}, (iii) update of the probabilities for processes at the wall~\cite{Booth_2020,Booth_2022} and (iv) revision of coefficients with typos in \cite{Annusova_2018}. The addition of $\mathrm{O_2(Hz)}$ is justified by its importance in the production of $\mathrm{O_2(a)}$ and in gas heating, as pointed out in \cite{Ionin_2007,Vasiljeva_2004,Braginskiy_2005} and detailed in sections \ref{sec:mainSpeciesProcesses} and \ref{sec:mainHeatProcesses}. The updates in the kinetic scheme concern only reactions for vibrationally-cold oxygen, i.e., involving the vibrational ground state of oxygen $\mathrm{O_2(X,0)}$, and are detailed in sections \ref{sec:eImpactV0} and \ref{sec:heavyV0}. The vibrational kinetics used in this work is entirely taken from~\cite{Annusova_2018} and is summarized in section \ref{sec:vibKin}. 

\subsubsection{Electron-impact processes for vibrationally-cold oxygen}\label{sec:eImpactV0}\quad\\
The electron-impact cross-sections of oxygen-containing plasmas used for the solution of the electron Boltzmann equation are presented and discussed by Alves~\textit{et~al.}~\cite{Alves_2016} and can be found in the IST-Lisbon database at LXCat~\cite{Oxyg_ISTLisbon}. The cross-section set for electron scattering with ground-state molecules was validated by comparing calculations with experimental data from swarm experiments. The set for $\mathrm{O_2}$ molecules includes the following mechanisms: effective (elastic and inelastic) momentum-transfer; excitations to rotational, vibrational and electronic levels; dissociation; dissociative attachment; and ionization. The set for O atoms includes an elastic momentum-transfer, excitations to electronic levels and ionization. For details on the cross sections used for solving the Boltzmann equation, the reader is referred to \cite{Alves_2016}, since in the present work we focus only on the electron-impact processes affecting the chemical kinetics.

Table \ref{tab:eImpactReactions} presents the electron-impact processes considered in the chemical kinetics. With the exception of R17, the cross sections for electron collisions with $\mathrm{O_2(X,0)}$ are originally taken from Phelps \cite{Phelps_1985_crossSections}. The cross sections for R3, R8 and R10 are based on the $R$-matrix calculations by Tashiro \textit{et al.} \cite{Tashiro_2006}. Due to lack of data, the cross sections for R9 and R11 are equal to the ones for the ground state, with the appropriate threshold shift. The collisions with $\mathrm{O(^3P)}$, R4 and R15, are according to Laher and Gilmore \cite{Laher_1990}. The cross section for electron-impact detachment of $\mathrm{O^-}$ (R16) is taken from the measurements of Vejby-Christensen~\textit{et al.}~\cite{Vejby-Christensen_1996}. The cross sections for the superelastic processes of R1-R5 are calculated through the Klein-Rosseland relation, expressing the principle of detailed balance \cite{Klein_1921}. The sum of the rate coefficients for dissociative recombination, R21 and R22, is according to the expression proposed by Gudmundsson and Lieberman~\cite{Gudmundsson_2004}, which is adequate for electron temperatures between 1 and 7 eV. The branching ratio 0.36:0.64 is based on the measurements of \cite{Petrignani_2005}, selecting the two main possible processes. Note that the value adopted for the branching is not critical, since the typical contribution for the production of $\mathrm{O(^1D)}$ is lower than $10^{-5}$. In the conditions of this study, R21 and R22 have a small role on the destruction of $\mathrm{O_2^+}$, lower than 2\%. 

% table with the electron-impact processes
{\input{oxygenElecImpactScheme.in}}

\subsubsection{Heavy-species processes for vibrationally-cold oxygen}\label{sec:heavyV0}\quad\\
Table \ref{tab:heavyReactions} presents the heavy-species processes taken into account in the kinetic scheme. The scheme is based on \cite{Annusova_2018}, with the following modifications. 

\begin{figure}[t]
	\centering
	\includegraphics[width=0.45\linewidth]{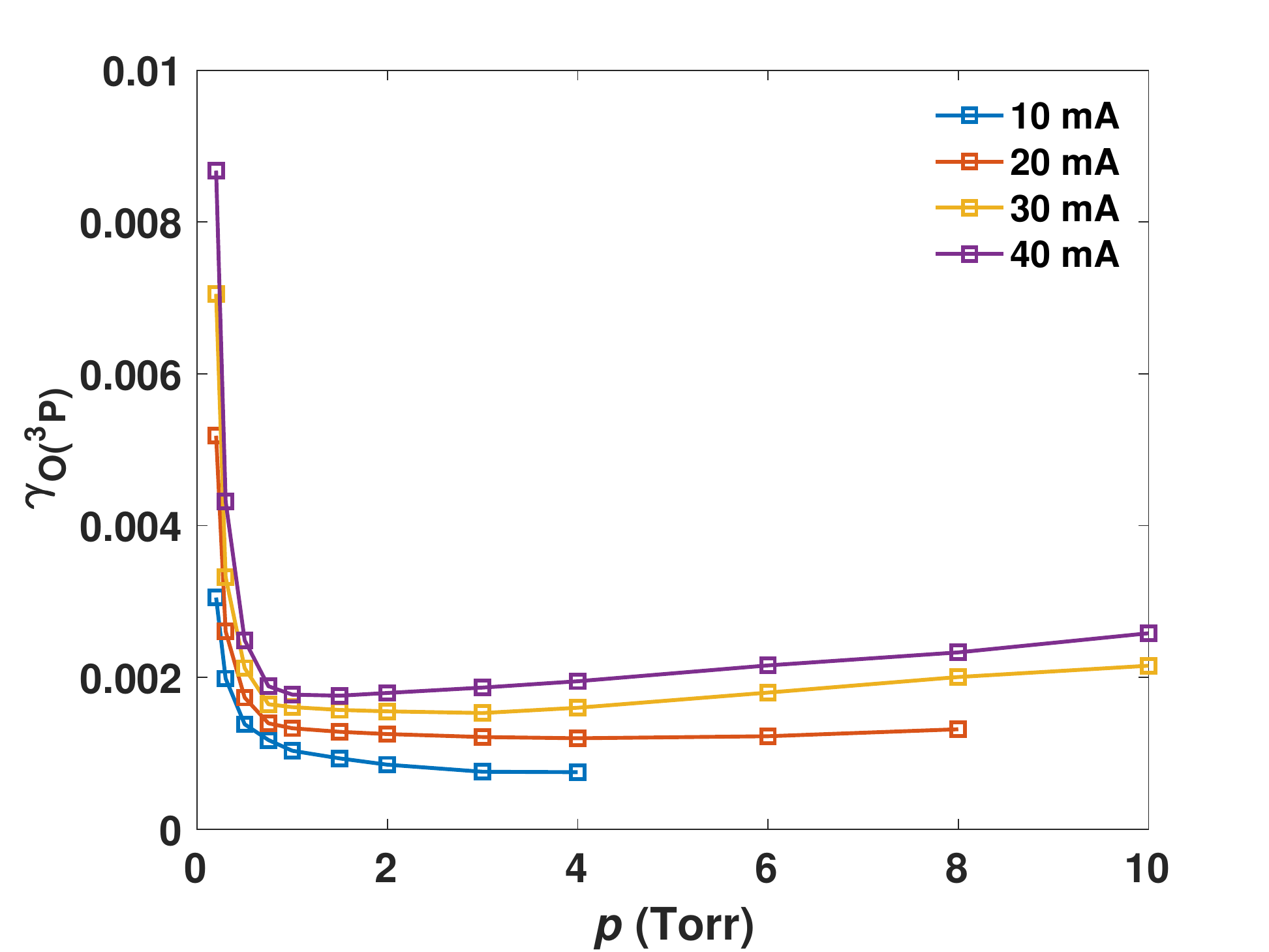}
	\caption{Experimental values for the $\mathrm{O(^3P)}$ wall recombination probability ($\gamma_\mathrm{O(^3P)}$) measured by Booth~\textit{et~al.}~\cite{Booth_2020,Booth_private}, as a function of pressure and various discharge currents, for a wall temperature $T\mathrm{w} = 50$ $\mathrm{^o}$C.}
	\label{fig:gammaO_Booth2020}
\end{figure}

\begin{itemize}

\item[-] The three-body quenching of $\mathrm{O_2(a)}$ (R24) proposed by Braginskiy~\textit{et~al.}~\cite{Braginskiy_2005} is now included in the kinetics, since it explains the behavior of $\mathrm{O_2(a)}$ density at higher pressures, as shown in section \ref{sec:validation}. Braginskiy~\textit{et~al.} estimate a rate coefficient for this reaction in the range $(1-3)\times10^{-44}$~$\mathrm{m^6s^{-1}}$ by means of an extensive comparison between model calculations and experimental measurements of $\mathrm{O_2(a)}$ and $\mathrm{O(^3P)}$. In this work, we opt by the highest value, $3\times10^{-44}$~$\mathrm{m^6s^{-1}}$, since it provides a remarkable agreement with experiment. Additionally, the probability of $\mathrm{O_2(a)}$ deactivation at the wall (R70) is updated from $5\times10^{-4}$ to $2.2\times10^{-4}$, following the recent measurements by Booth~\textit{et~al.}~\cite{Booth_2020}.

\item[-] The quenching of $\mathrm{O_2(b)}$ with $\mathrm{O(^3P)}$ (R28) is updated according to the work by Booth~\textit{et~al.}~\cite{Booth_2022}. The coefficient is now composed by the constant value proposed by Ionin~\textit{et~al.}~\cite{Ionin_2007} together with the $T_\mathrm{g}$-dependent reactive component proposed in \cite{Booth_2022}. The latter component is based on a very complete comparison between modelling and experiment of both average and radial profiles of $\mathrm{O_2(b)}$ densities, at 0.3--10 Torr gas pressure and 10--40 mA current. Furthermore, reaction R29 is added with a coefficient consistent with its inverse reaction, R43. As shown in section \ref{sec:mainSpeciesProcesses}, this reaction starts to be important at higher pressures. Lastly, the probability of $\mathrm{O_2(b)}$ deactivation at the wall (R71) is updated from 0.02 to 0.135, according to the measurements by Booth~\textit{et~al.}~\cite{Booth_2022}.

\item[-] The quenching kinetics (R30-R35) and the neutral transport (R72) of $\mathrm{O_2(Hz)}$ are added in accordance with Vasiljeva~\textit{et~al.}\cite{Vasiljeva_2004}.

\item[-] The three-body recombination of $\mathrm{O(^3P)}$ (R40) is now consistent with Gordiets~\textit{et~al.}~\cite{Gordiets_1995}. However, this process is not very relevant for the studied conditions. The $\mathrm{O(^3P)}$ wall recombination probability, $\gamma_\mathrm{O(^3P)}$, strongly depends on the wall temperature, gas pressure and discharge current. We take the values of $\gamma_\mathrm{O(^3P)}$ measured for the corresponding conditions by Booth~\textit{et~al.}~\cite{Booth_2020,Booth_private}, represented in figure~\ref{fig:gammaO_Booth2020}. Note that the same group measured $\gamma_\mathrm{O(^3P)}$ for the same conditions in a previous work~\cite{Booth_2019}, but those values are systematically lower than in~\cite{Booth_2020}. Work is in progress to self-consistently calculate $\gamma_\mathrm{O(^3P)}$ by coupling both the surface and volume kinetics~\cite{Marinov_2017}. This may help to disentangle the reasons for the differences between the two sets of experimental measurements.

\item[-] Lastly, as referred in section~\ref{sec:LoKIC}, the loss of positive ions (R68 and R69) is described by ambipolar diffusion, with a correction in the diffusion length induced by the presence of the $\mathrm{O^-}$ ion, as described in~\cite{Guerra_1999}. The loss of negative ions at the wall is not considered, as they are mostly confined to the volume and virtually none reach the wall~\cite{Ferreira_1988,Daniels_1990,Guerra_1999}.
\end{itemize}
% table with the heavy-species processes
\input{oxygenHeavySpeciesScheme.in}

\subsubsection{Vibrational kinetics of $O_2(X,0:41)$}\label{sec:vibKin}\quad\\
The kinetics of vibrationally-excited $\mathrm{O_2(X,0:41)}$ molecules is very well detailed and validated in \cite{Annusova_2018}. The vibrational kinetics contains electron-impact (e-V), vibration-to-translation (V-T), vibration-to-vibration (V-V) and wall de-excitation processes, as shown in table \ref{tab:vibKinetics} and summarized in the following paragraphs.

%  table with the vibrational kinetics
{\input{oxygenVibrationsScheme.in}}

The electron-impact cross-sections for the excitations of $\mathrm{O_2(a)}$ and $\mathrm{O_2(b)}$ from $\mathrm{O_2(X,v>0)}$, R76 and R77, were calculated by Annu\v{s}ova~\textit{et~al.}~\cite{Annusova_2018}, using a scaling of the $\mathrm{O_2(X,0)}$ cross section according to the corresponding Franck-Condon factors, similarly to what was done by Loureiro~\textit{et~al.}~\cite{Loureiro_1990} in nitrogen discharges. The electron-impact cross-sections for vibrational excitation (R78) are taken from the work by Laporta~\textit{et~al.}~\cite{Laporta_2013}, where the $R$-matrix method was used to calculate resonant interactions between the 42 bound v-levels. The cross sections for electron-impact dissociation leading to $\mathrm{O(^3P)}$ (R79) and for dissociative attachment (R82) are scaled from the vibrational ground state to higher levels~\cite{Annusova_2018}, according to the trends of the theoretical calculations made by Laporta~\textit{et~al.}~\cite{Laporta_2015}. The cross sections for electron-impact dissociation leading to $\mathrm{O(^1D)}$ (R80) and ionization (R81) are assumed to be the same for all vibrational levels and equal to the ones used for $\mathrm{O_2(X,0)}$.

The rate coefficients for V-T $\mathrm{O_2-O}$ processes (R83) are obtained from the quasiclassical trajectory calculations of Esposito~\textit{et~al.}~\cite{Esposito_2008}. These computations were fitted to a formula depending on several coefficients, allowing to easily calculate the rate coefficients for multi-quanta transitions $\mathrm{\Delta v = 1 : 30}$ in the temperature range $T_\mathrm{g} = 50 - 10000$ K~\cite{Esposito_2008}. 

The V-T and V-V $\mathrm{O_2-O_2}$ rate coefficients were computed using the forced harmonic oscillator (FHO) model~\cite{Adamovich_1998}, initially performed by Lino da Silva~\textit{et~al.}~\cite{LinodaSilva_2012} and later revisited by Annu\v{s}ova~\textit{et~al.}~\cite{Annusova_2018}. In the latter work, the FHO calculations were redone for 42 bound vibrational levels, instead of 46. Moreover, the rates for vibrational transitions whose energy difference is comparable with the gas thermal energy were corrected, so as to compensate the limitations of the FHO approach in those conditions~\cite{LinodaSilva_2012}.
The final V-T and V-V rate-coefficients were fitted as a function of the vibrational quantum numbers and $T_\mathrm{g}$, applicable in the temperature range $T_\mathrm{g} = 300-1000$ K~\cite{Annusova_2018}.

Finally, the vibrational kinetics includes wall de-excitation of $\mathrm{O_2(X,1:41)}$, assuming a deactivation probability equal to the one measured for nitrogen~\cite{Marinov_2012}, due to lack of information concerning this process in oxygen.

\subsection{Transport and thermodynamic data}\label{sec:transpThermoData}

In addition to a comprehensive description of the reaction mechanism, detailed knowledge of physical parameters, e.g. mobility, diffusion coefficient, thermal conductivity and enthalpy, is required to obtain meaningful modelling results.

The transport of neutral species to the wall (cf. section \ref{sec:LoKIC}) requires the species diffusion coefficients, $D_n$.
In this work, $D_n$ is given by Wilke's formula for multicomponent mixtures \cite{Cheng_2008}:
\begin{equation}
D_n = \frac{1-x_n}{\sum\limits_{\forall l \neq n} x_l/D_{n,l}}\ ,
\end{equation}
where $D_{n,l}$ is the binary diffusion coefficient of species $n$ in species $l$ and $x_{n} = N_n/N$ is the relative concentration (molar fraction) of species $n$. The binary diffusion coefficients $D_{n,l}$ can be calculated as in Hirschfelder~\textit{et~al.} \cite{Hirschfelder_1964} (section 8.2, equation 8.2-44) and Guerra~\textit{et~al.}\cite{Guerra_2019} (section 4.1):
\begin{equation}
D_{n,l} [\mathrm{m^2 s^{-1}}] = \frac{1.929\times 10^{21} \sqrt{T_\mathrm{g}/(2\mu_{nl})}}{N\sigma_{nl}^2[\mathrm{\AA^2}]\Omega^{(1,1)*}k_\mathrm{B}T_\mathrm{g}/\epsilon_{nl}} \ ,
\end{equation}
where $\mu_{nl}$ is the reduced mass, $T_\mathrm{g}$ is the gas temperature, $\sigma_{nl} = (\sigma_n + \sigma_l)/2$ and $\epsilon_{nl} = \sqrt{\epsilon_n\epsilon_l}$ are the Lennard-Jones binary interaction potential parameters, and the collision integral $\Omega^{(1,1)*}$ is tabulated in \cite{Hirschfelder_1964}. The Lennard-Jones parameters, $\epsilon_n$ and $\sigma_n$, of a species $n$ are assumed to be independent of the electronic and vibrational state, and the corresponding values for O, O$_2$ and O$_3$ species are taken from \cite{CHEMKIN_2000} and presented in table \ref{tab:LennardJonesParameters}.

\begin{figure}[t]
  \centering
  \begin{minipage}[H]{0.49\linewidth}
  \centering
     \includegraphics[width=0.9\linewidth]{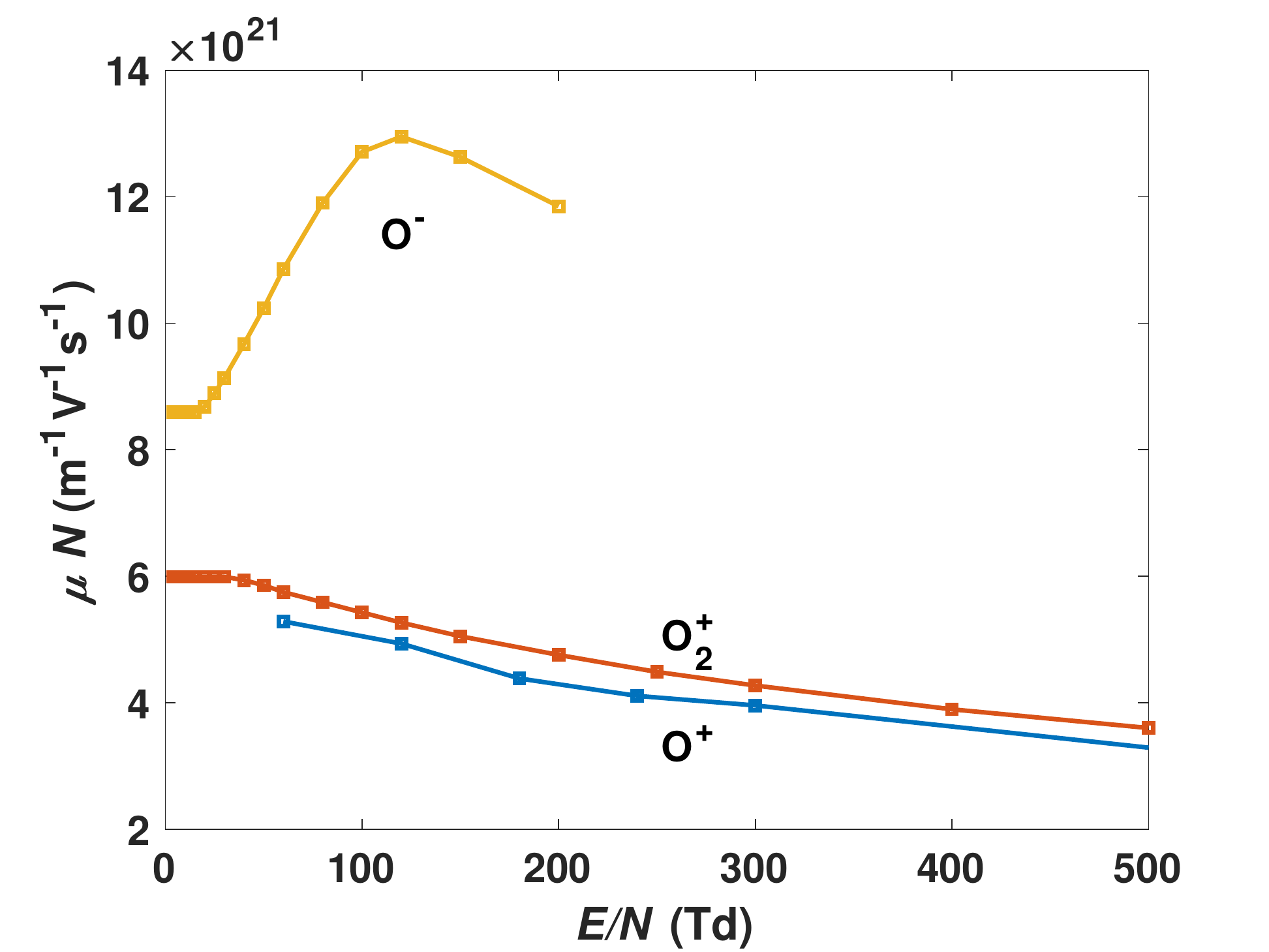}
    \caption{Ion reduced mobilities, $\mu_i N$, as a function of $E/N$. See text for references.}
    \label{fig:ionRedMob}
  \end{minipage}  
  \hfill
  \begin{minipage}[H]{0.49\linewidth}
  \centering
     \includegraphics[width=0.9\linewidth]{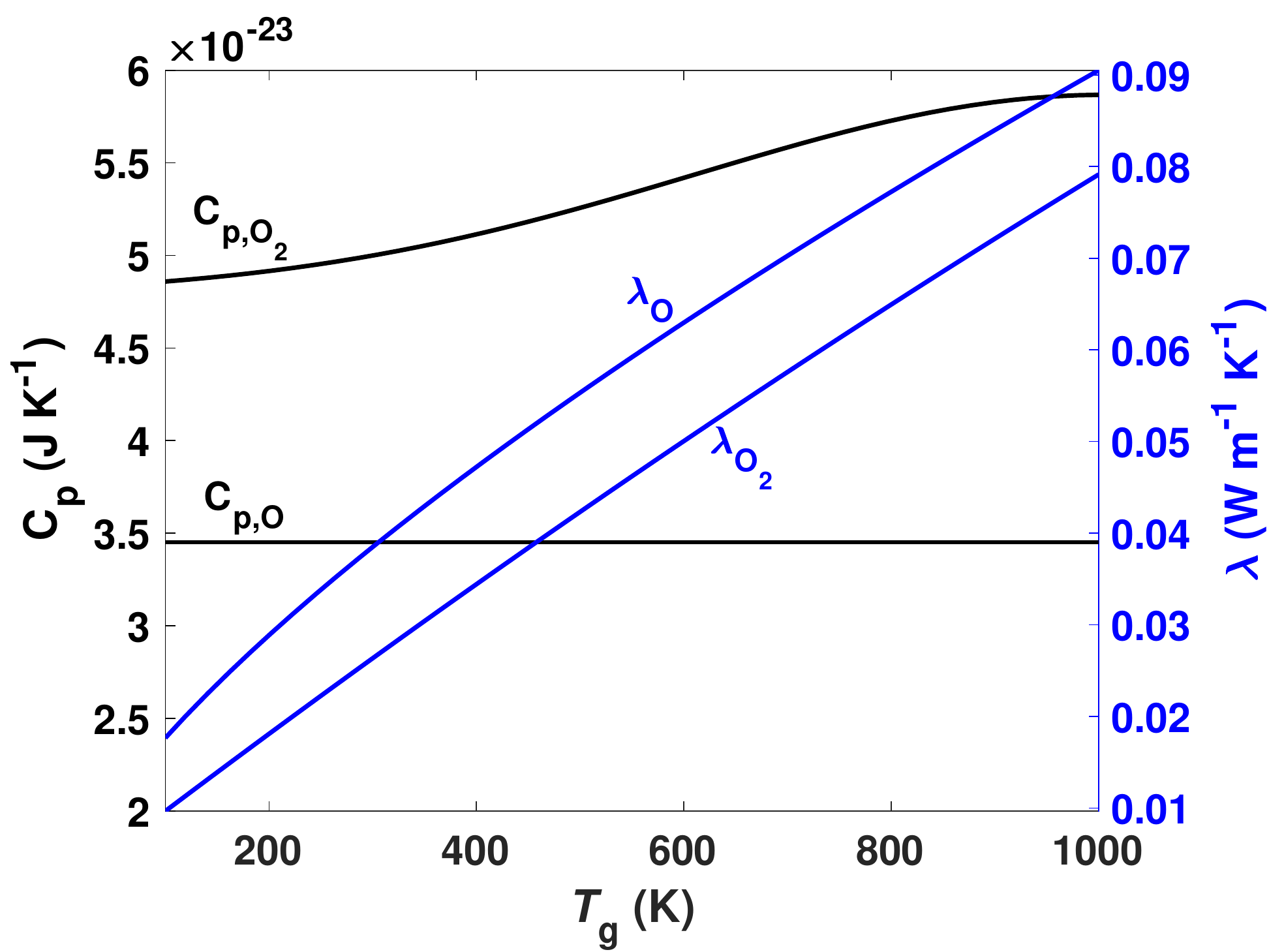}
    \caption{Heat capacities at constant pressure ($C_p$, left) and thermal conductivities ($\lambda$, right) of $\mathrm{O_2}$ and O. See text for references. }
    \label{fig:capacityConductivity}
  \end{minipage}  
\end{figure}

The mobility and diffusion coefficients of charged species are necessary for the calculation of ambipolar transport~\cite{Guerra_1999}. The  ion reduced mobilities, $\mu_i N$, of $\mathrm{O_2^+}$\cite{Ellis_1976}, $\mathrm{O^+}$\cite{Eliasson_1986} and $\mathrm{O^-}$\cite{Ellis_1976} are taken from experimental values reported in the literature, for background O$_2$ molecules, with a dependence on $E/N$, as shown in figure~\ref{fig:ionRedMob}. The reduced mobility of $\mathrm{O^+}$ is not available for $E/N < 60$~Td, and in this interval is taken with the value at $E/N = 60$~Td. The ion reduced diffusion coefficients, $D_i N$, are deduced from the Einstein relation, $D_i/\mu_i = k_B T_i$, with the ion temperature $T_i$ assumed equal to the gas temperature $T_\mathrm{g}$. The electron mobility and diffusion coefficients are taken from the solution of the electron Boltzmann equation calculated by LoKI-B.

The heat capacities and the thermal conductivities of oxygen species are needed for the self-consistent calculation of the gas temperature. In turn, these quantities have an important dependence with gas temperature, as evidenced in figure~\ref{fig:capacityConductivity}. The heat capacity of $\mathrm{O_2}$ at constant pressure $(C_{p,\mathrm{O_2}}$) is taken from~\cite{Pintassilgo_2014,Younglove_1982}. Assuming atomic oxygen as an ideal gas, the heat capacity has a constant value $C_{p,\mathrm{O}} = 5/2 k_B$. The heat capacity of $\mathrm{O_3}$ is assumed to have a negligible impact on the results, due to the low $\mathrm{O_3}$ concentration in the conditions under study. Therefore, the heat capacity of the mixture is estimated using $C_p = (1-\frac{[O]}{N}) C_{p,\mathrm{O_2}} + \frac{[O]}{N}C_{p,\mathrm{O}}$, where $\frac{[O]}{N}$ is the O-atom density fraction. The thermal conductivities ($\lambda$) of $\mathrm{O_2}$ and O are taken from~\cite{Pintassilgo_2014,Hanley_1973} and~\cite{Dalgarno_1962}, respectively. The thermal conductivity of the multicomponent mixture is calculated as presented by Mason and Saxena~\cite{Mason_1958}. Note that for the steady-state calculations presented in this work, the model results are sensitive to the thermal conductivity but not to the heat capacity, as can be understood from equation~(\ref{eq:heatTransferAverage}).

Lastly, the enthalpies of the oxygen species are necessary to calculate the energies released for gas heating ($\epsilon_\mathrm{gh}$) in the various reactions. The values of $\epsilon_\mathrm{gh}$ are presented in the third columns of tables \ref{tab:eImpactReactions}, \ref{tab:heavyReactions} and \ref{tab:vibKinetics}. The enthalpies of formation for the species considered in this work are presented in table \ref{tab:stateEnergies}. Note that the energies of the $\mathrm{O_2(X,v)}$ vibrational levels are taken from Laporta~\textit{et~al.}~\cite{Laporta_2013} and that the $\mathrm{O_3^*}$ enthalpy corresponds to the first vibrational level of ozone~\cite{Lopaev_2011,Marinov_2013}. For reactions between heavy particles, $\epsilon_\mathrm{gh}$ is given by the difference between the summed enthalpies of formation for the reactants and the products. For reactions occurring at the wall, the procedure is similar but the energy release is pondered by the energy fraction transferred to the volume ($f_\mathrm{wall}=0.5$)\cite{Pintassilgo_2014}. Concerning electron-impact processes, only the ones involving dissociation contribute to gas heating. In this case, $\epsilon_\mathrm{gh}$ is calculated as the difference between the enthalpy of formation of the dissociative state and the summed enthalpies of the dissociation products. The enthalpy of formation of the dissociative state is taken as the sum of the heavy-target enthalpy and the energy threshold of the electron scattering cross section for the corresponding process.

% \usepackage{booktabs}
\begin{table}[t]
\centering
\caption{Lennard-Jones parameters ($\epsilon_n$ and $\sigma_n$) used in this work for oxygen species, taken from \cite{CHEMKIN_2000}.}
\label{tab:LennardJonesParameters}
\begin{tabular}{@{}lll@{}}
\toprule
Species & $\epsilon_n/k_\mathrm{B}$ (K) & $\sigma_n (\mathrm{\AA})$ \\ \midrule
O       & 80.0                          & 2.75             \\
O$_2$   & 107.4                         & 3.458            \\
O$_3$   & 180.0                         & 4.1              \\ \bottomrule
\end{tabular}
\end{table}

% \usepackage{booktabs}
\begin{table}[t]
\centering
\scriptsize
\caption{Enthalpies of the states considered in this work.}
\label{tab:stateEnergies}
\begin{tabular}{@{}lll|lll@{}}
\toprule
Species            & Enthalpy (eV) & Ref.                  & Species          & Enthalpy (eV) & Ref.                 \\ \midrule
$\mathrm{O_2(X,v)}$ & $E_\mathrm{v}$ & \cite{Laporta_2013}  & $\mathrm{O_3}$ & 1.48 & \cite{Eliasson_1986}                          \\
$\mathrm{O_2(a)}$   & 0.98           & \cite{Eliasson_1986} & $\mathrm{O_3^*}$  & 1.57 & \cite{Eliasson_1986,Lopaev_2011,Marinov_2013} \\
$\mathrm{O_2(b)}$  & 1.63        & \cite{Eliasson_1986}  & $\mathrm{O_2^+}$ & 12.14       & \cite{Eliasson_1986} \\
$\mathrm{O_2(Hz)}$ & 4.5         & \cite{Vasiljeva_2004} & $\mathrm{O^+}$   & 16.26       & \cite{Eliasson_1986} \\
$\mathrm{O(^3P)}$  & 2.58        & \cite{Eliasson_1986}  & $\mathrm{O^-}$   & 1.05        & \cite{Eliasson_1986} \\
$\mathrm{O(^1D)}$  & 4.55        & \cite{Eliasson_1986}  &                  &             &                      \\ \bottomrule
\end{tabular}
\end{table}

\pagebreak

\section{Results}\label{sec:results}
\subsection{Validation}\label{sec:validation}
In this section, we present the results that allow to confirm the reaction mechanism for $\mathrm{O_2}$ plasmas, by comparison of the model predictions with the experiment for gas pressures between 0.2 and 10 Torr. We start by validating the self-consistent reduced electric field and gas temperature (average values and radial profile), and then focus on the comparison of the main species densities. Since the effect of gas pressure variation is rather similar for all discharge currents, we present the results for 30 mA, where more experimental data are available. However, we also study the impact of discharge current on the results, for gas pressures of 0.5 and 3 Torr. 

\subsubsection{Reduced electric field and gas temperature}
\begin{figure}[H]
  \centering
  \begin{minipage}[H]{0.46\linewidth}
  \centering
     \includegraphics[width=0.92\linewidth]{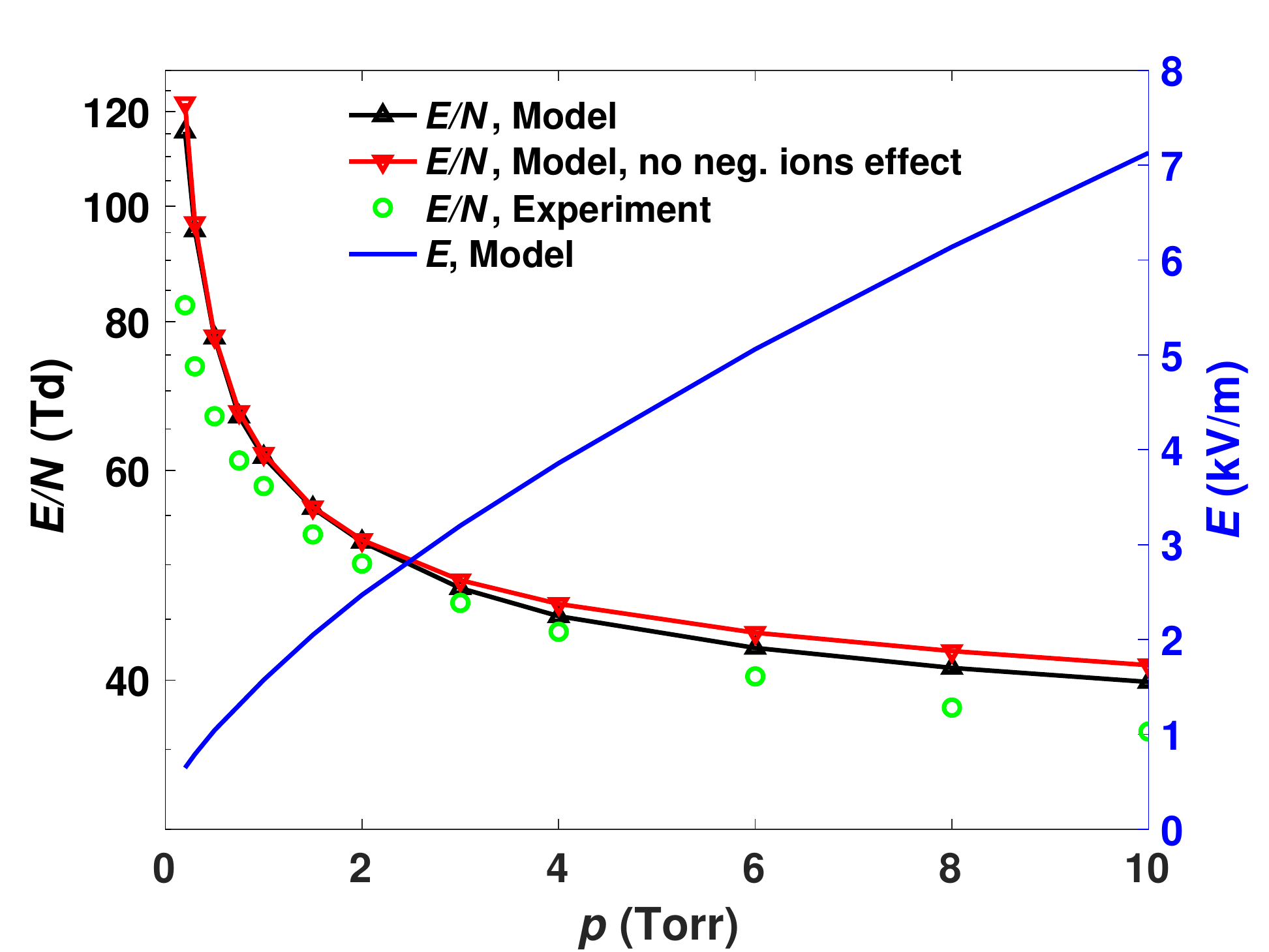}
    \subcaption{}
    \label{fig:EN30mA}
  \end{minipage}
  \hfill
  \begin{minipage}[H]{0.46\linewidth}
  \centering
     \includegraphics[width=0.92\linewidth]{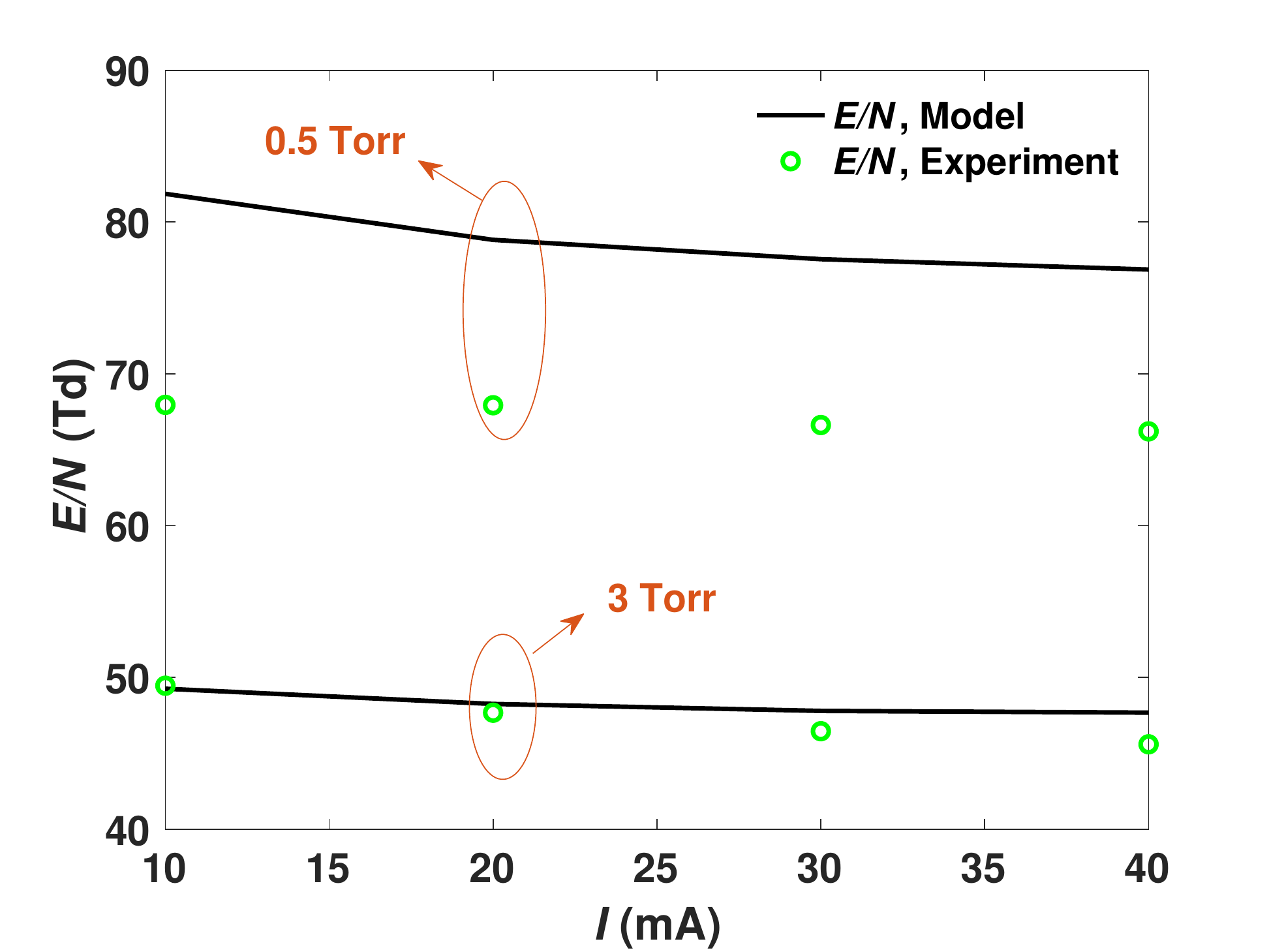}
    \subcaption{}
    \label{fig:ENVarCurr}
  \end{minipage}
  \label{fig:EN}
  \caption{Comparison between modelling and experimental\cite{Booth_2019,Booth_private} values of the reduced electric field as a function of (a) the gas pressure, for a discharge current of 30 mA, and (b) the discharge current, for gas pressures of 0.5 and 3 Torr. In figure (a), the electric field is also plotted as a function of pressure.}
  \label{fig:EN}
\end{figure}

\begin{figure}[H]
	\centering
	\includegraphics[width=0.42\linewidth]{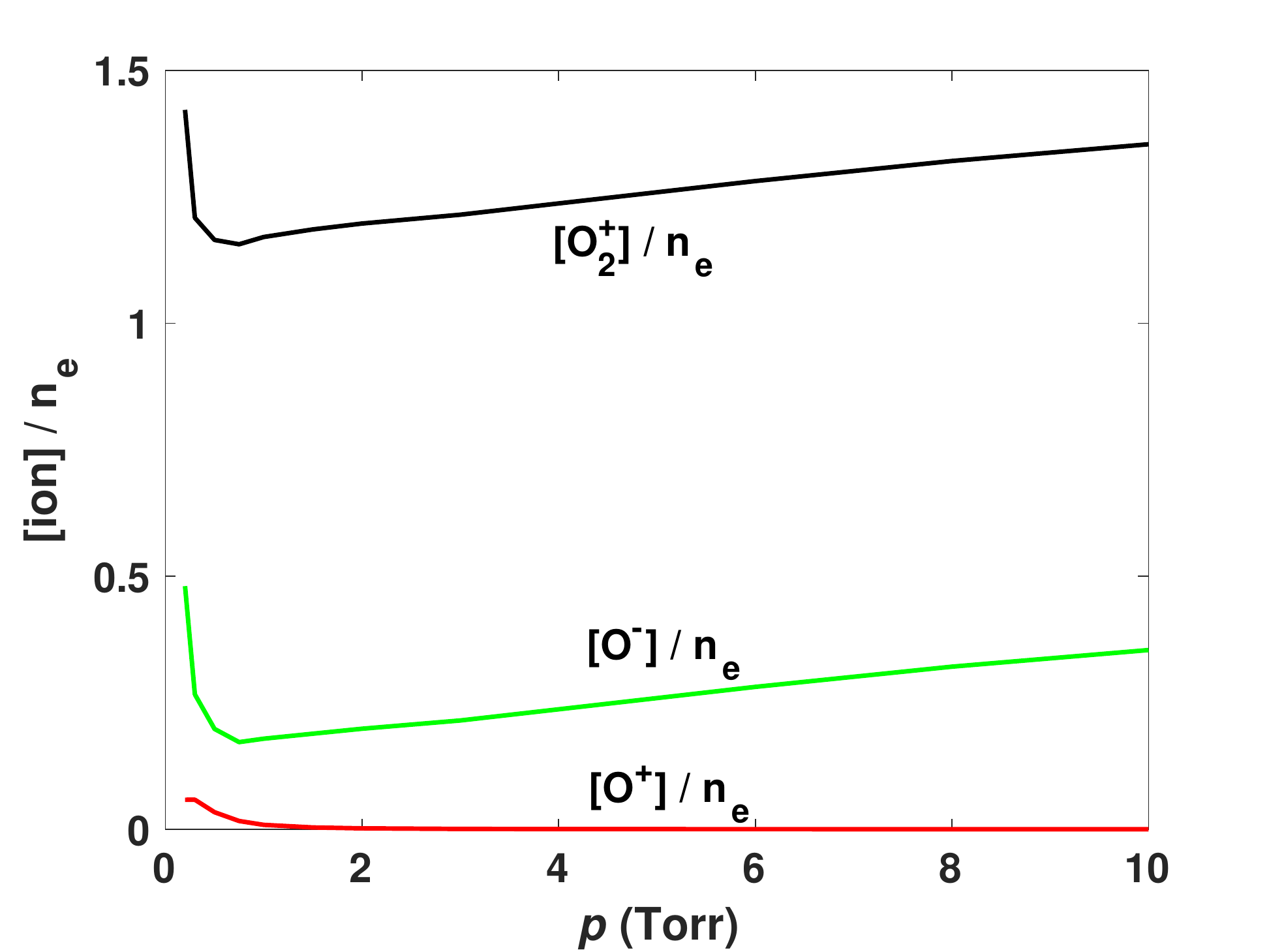}
	\caption{Ion densities relative to the electron density as a function of gas pressure, for a discharge current of 30 mA.}
	\label{fig:ions30mA}
\end{figure}

\quad\\
Figure~\ref{fig:EN} shows the comparison between modelling and experimental~\cite{Booth_2019,Booth_private} values of the reduced electric field $E/N$. The model captures rather well the tendency with pressure and predicts values in agreement with those found in experiment (see figure~\ref{fig:EN30mA}). However, for pressures below 1 Torr, the model overestimates $E/N$, with an increasing effect with decreasing pressure. This might be related with the approximations made in assuming ambipolar diffusion, i.e. high collisionality regime, which start to fail in this pressure range. An analysis of the influence on the results of adopting different charged-particle transport models is presented in \cite{Alves_2023}. Moreover, when comparing to model results considering classical ambipolar diffusion (i.e. without the influence of negative ions), we find that the $\mathrm{O^-}$ ions lead to a small decrease in $E/N$, associated to a decrease in the loss of positive ions at the wall. The correction induced by $\mathrm{O^-}$ ions is proportional to its content relatively to the electron density, as can be verified by inspection of figure~\ref{fig:ions30mA}. For this reason, the difference between the models is more noticeable for pressures below 0.5 Torr and above 2 Torr. From figure \ref{fig:ENVarCurr}, we find that the model predicts a small decrease in $E/N$ with increasing current, consistent with what is observed in experiment. A larger current / electron density leads to higher densities of reactive species, such as $\mathrm{O(^3P)}$ and $\mathrm{O_2(a)}$. The presence of these species causes an increase in the high-energy tail of the EEDF, associated with weaker inelastic collisions and stronger superelastic collisions, due to $\mathrm{O(^3P)}$ and $\mathrm{O_2(a)}$, respectively. Therefore, for a given $E/N$, the ionization coefficient is higher in both cases, originating a lower maintenance electric field with increasing current.

\begin{figure}[H]
  \centering
  \begin{minipage}[H]{0.46\linewidth}
  \centering
     \includegraphics[width=0.95\linewidth]{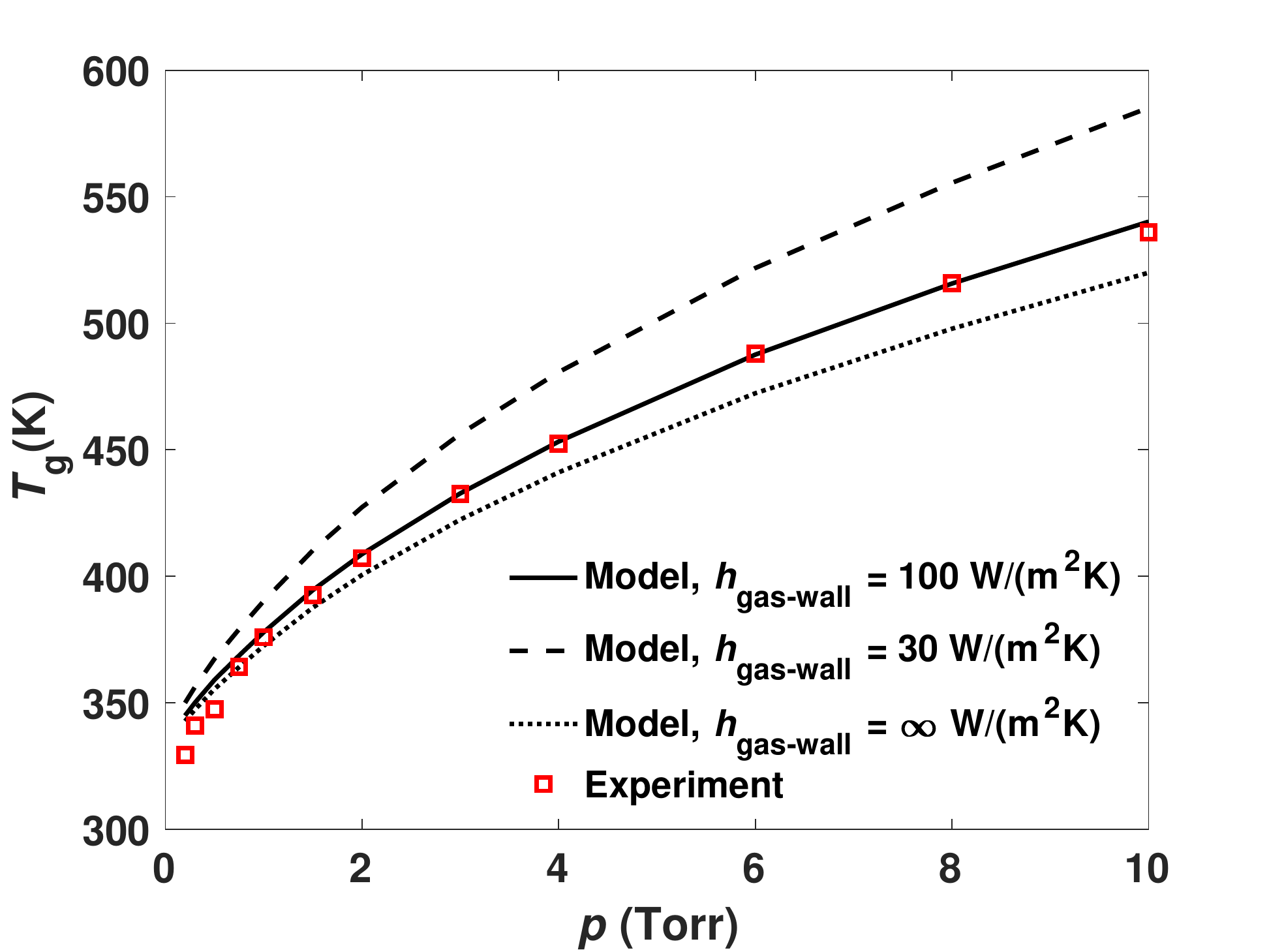}
    \subcaption{}
    \label{fig:TgasAver30mA}
  \end{minipage}
  \hfill
  \begin{minipage}[H]{0.46\linewidth}
  \centering
     \includegraphics[width=0.95\linewidth]{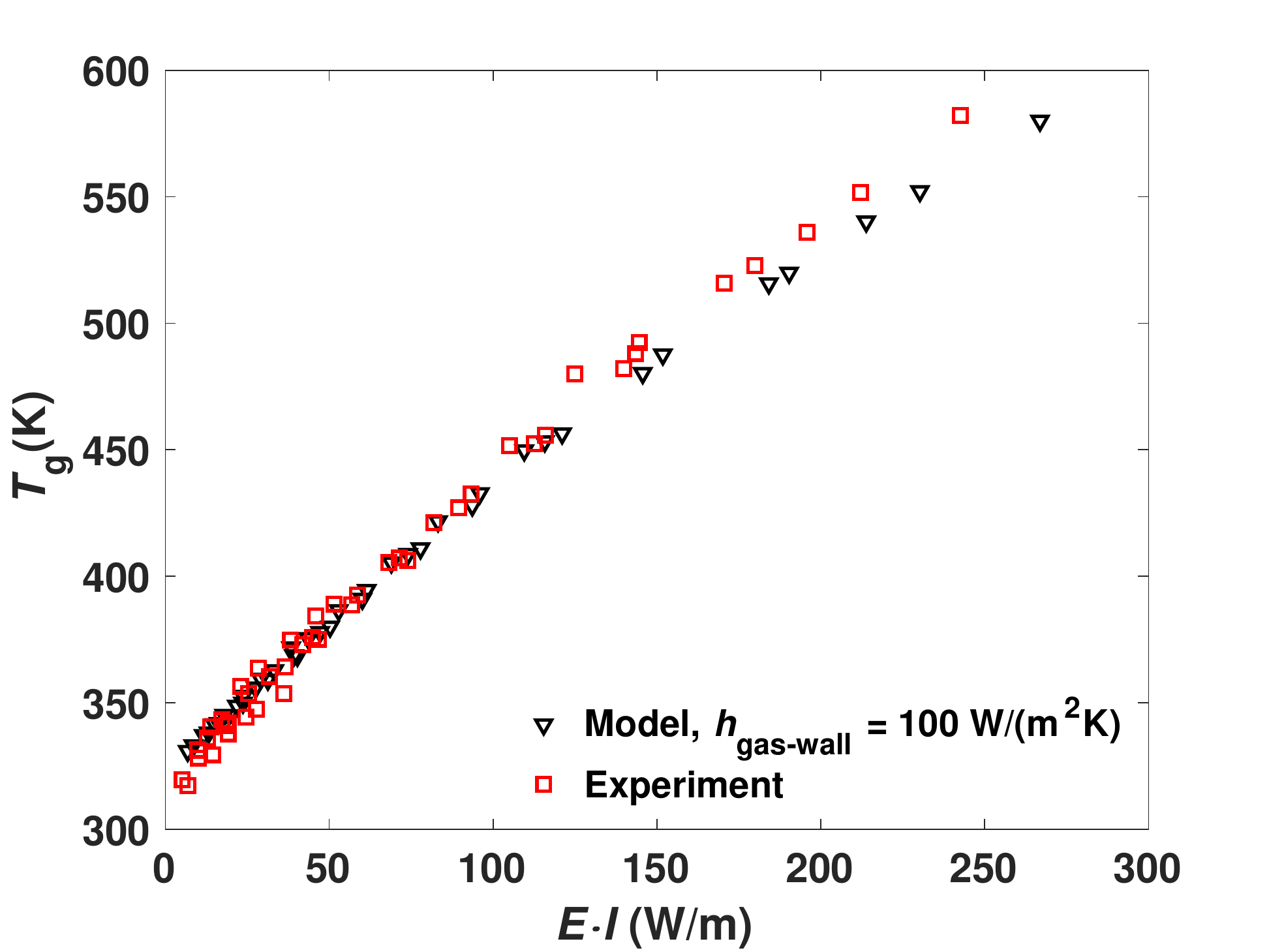}
    \subcaption{}
    \label{fig:TgasAverAllCurrents}
  \end{minipage}
  \caption{Comparison between modelling and experimental\cite{Booth_2019,Booth_private} results of the average gas temperature ($T_\mathrm{g}$) as a function of (a) the gas pressure, for a discharge current of 30 mA, and (b) the discharge power per unit length ($E\cdot I$), for all pressures and 10-40 mA discharge currents.}
  \label{fig:TgasAver}
\end{figure}

The average values of gas temperature ($T_\mathrm{g}$) are compared against experiment~\cite{Booth_2019,Booth_private} in figure~\ref{fig:TgasAver}. In figure \ref{fig:TgasAver30mA}, we see that a convection coefficient between the gas and the wall~($h_\mathrm{gas-wall}$) of 100 $\mathrm{W/(m^2 K)}$ provides a remarkable agreement with experiment. For this reason, we adopt this value in all simulations. Note that if we use $h_\mathrm{gas-wall} = 30\ \mathrm{W/(m^2 K)}$, the model visibly overestimates $T_\mathrm{g}$, whereas if we use $h_\mathrm{gas-wall} = \infty \Leftrightarrow T_\mathrm{nw} = T_\mathrm{w}$ (see also equation \ref{eq:Tnw}), the model underestimates the $T_\mathrm{g}$ values for pressures higher than 1 Torr, albeit the correct trend is captured in all cases. In figure~\ref{fig:TgasAverAllCurrents}, $T_\mathrm{g}$ is plotted against the discharge power per unit length ($E\cdot I$). This figure evidences that $T_\mathrm{g}$ is nearly proportional to $E\cdot I$. At high $E\cdot I$, the deviation of model results of $T_\mathrm{g}(E\cdot I)$ from the experimental trend apparent in the figure is mainly caused by the slight overestimation of $E/N$ values predicted by the model at high pressures, which influences directly the values in the horizontal axis, $E\cdot I$. Nevertheless, we should underline that for each set $(p,I)$ the agreement between the model predictions for $T_\mathrm{g}$ and experiment is very good.  

\begin{figure}[H]
	\centering
	\vspace{-0.3cm}
	\includegraphics[width=0.55\linewidth]{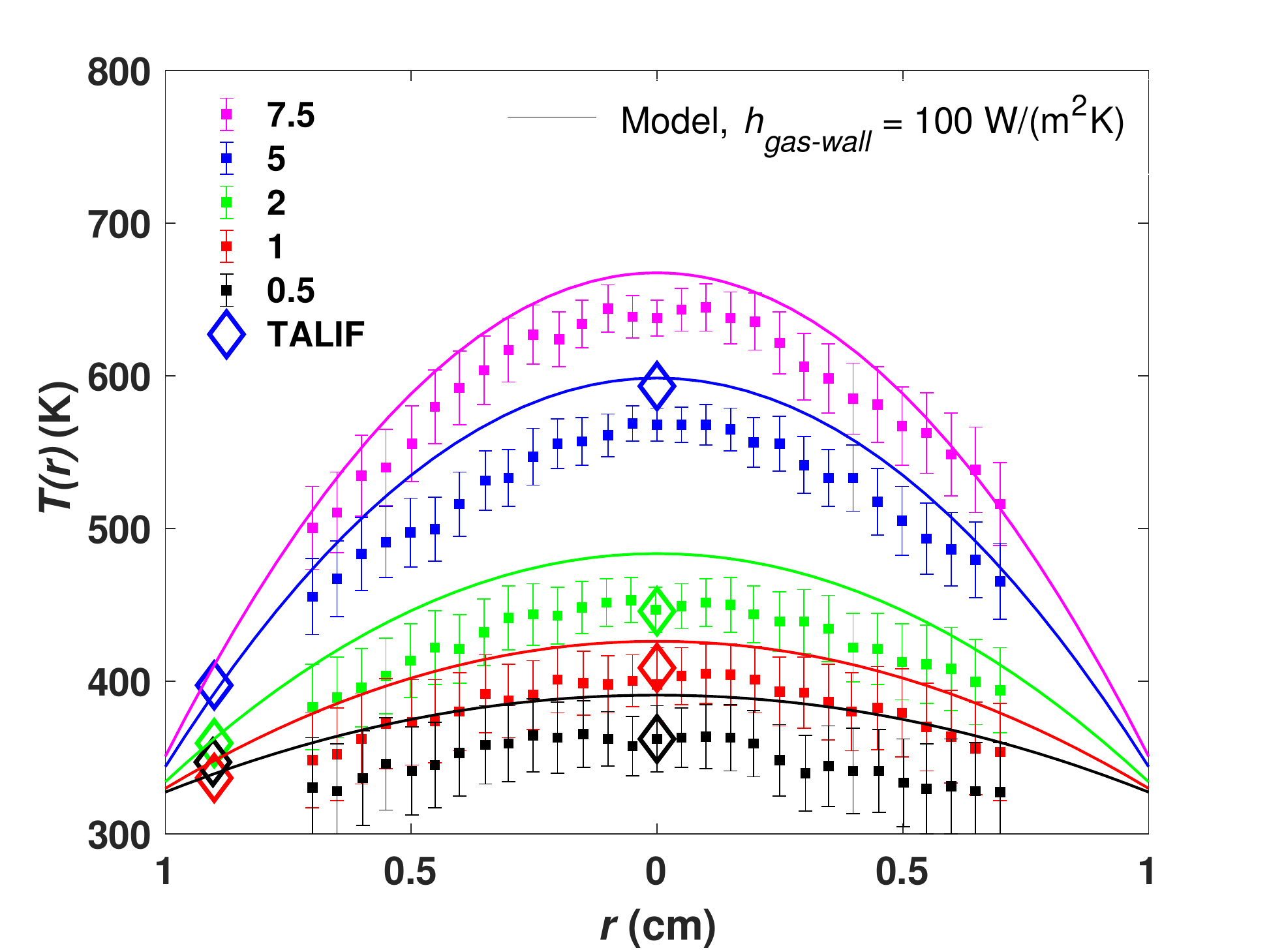}
	\caption{Comparison between modelling and experimental\cite{Booth_2019} values of the radial profile of gas temperature ($T(r)$) for different pressures and a discharge current of 30 mA. The squares and diamonds are experimental measurements from $\mathrm{O_2(b)}$ emission and TALIF, respectively\cite{Booth_2019}.}
	\label{fig:TgasRadial30mA}
\end{figure}

Note that, although the model is 0D, i.e. spatially averaged, it is possible to reconstruct the radial profile assumed in equation~(\ref{eq:Tradial}). The calculated profile of gas temperature is compared against experiment for a discharge current of 30 mA and different pressures, in figure \ref{fig:TgasRadial30mA}. Despite the significant approximations made in the calculation of gas temperature (cf. section \ref{sec:LoKIC}), the accordance between the model and the measurements is quite good, and it further confirms the choice for the value of the convection coefficient. Notice that the radial profiles from the model are always slightly higher than in the experiment, while the average values of gas temperature reported in figure~\ref{fig:TgasAver30mA} are not. This seemingly contradictory result may stem from the different measuring set-ups used for the determination of these temperatures. While the radially-averaged measurements are performed across the tube, the radially-resolved measurements are line-of-sight (for more details, see figure 1 and section 2.3 of \cite{Booth_2019}). Therefore, the experimental data reported in figure~\ref{fig:TgasAver30mA} are obtained independently from the measurements shown in figure~\ref{fig:TgasAverAllCurrents}, and the former are not a simple radial average of the latter (albeit not far from that).

\subsubsection{Main species}
\begin{figure}[H]
  \centering
  \vspace{-0.2cm}
  \begin{minipage}[H]{0.46\linewidth}
  \centering
     \includegraphics[width=0.95\linewidth]{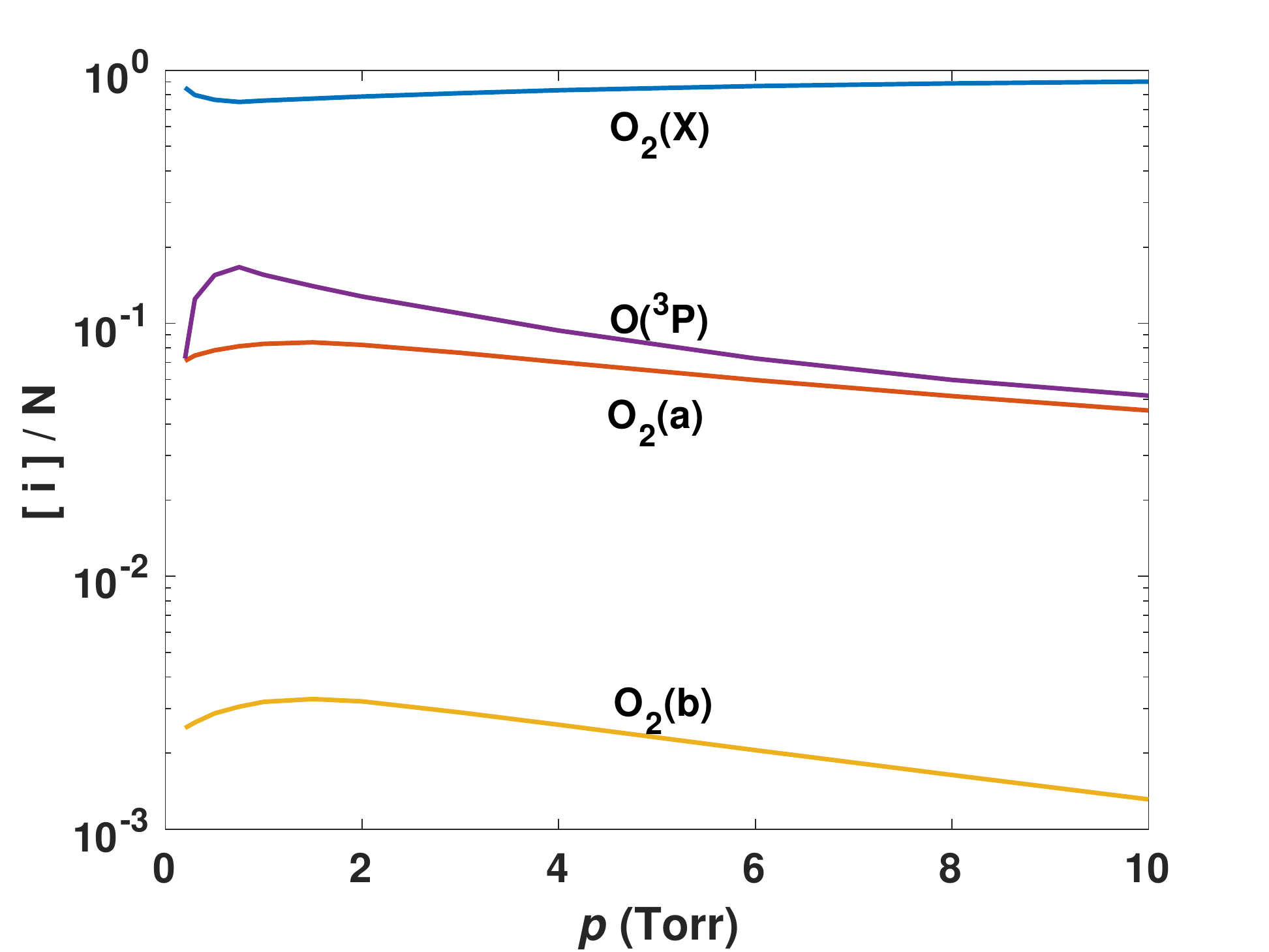}
    \subcaption{}
    \label{fig:MajorSpecies30mA}
  \end{minipage}
  \hfill
  \begin{minipage}[H]{0.46\linewidth}
  \centering
     \includegraphics[width=0.95\linewidth]{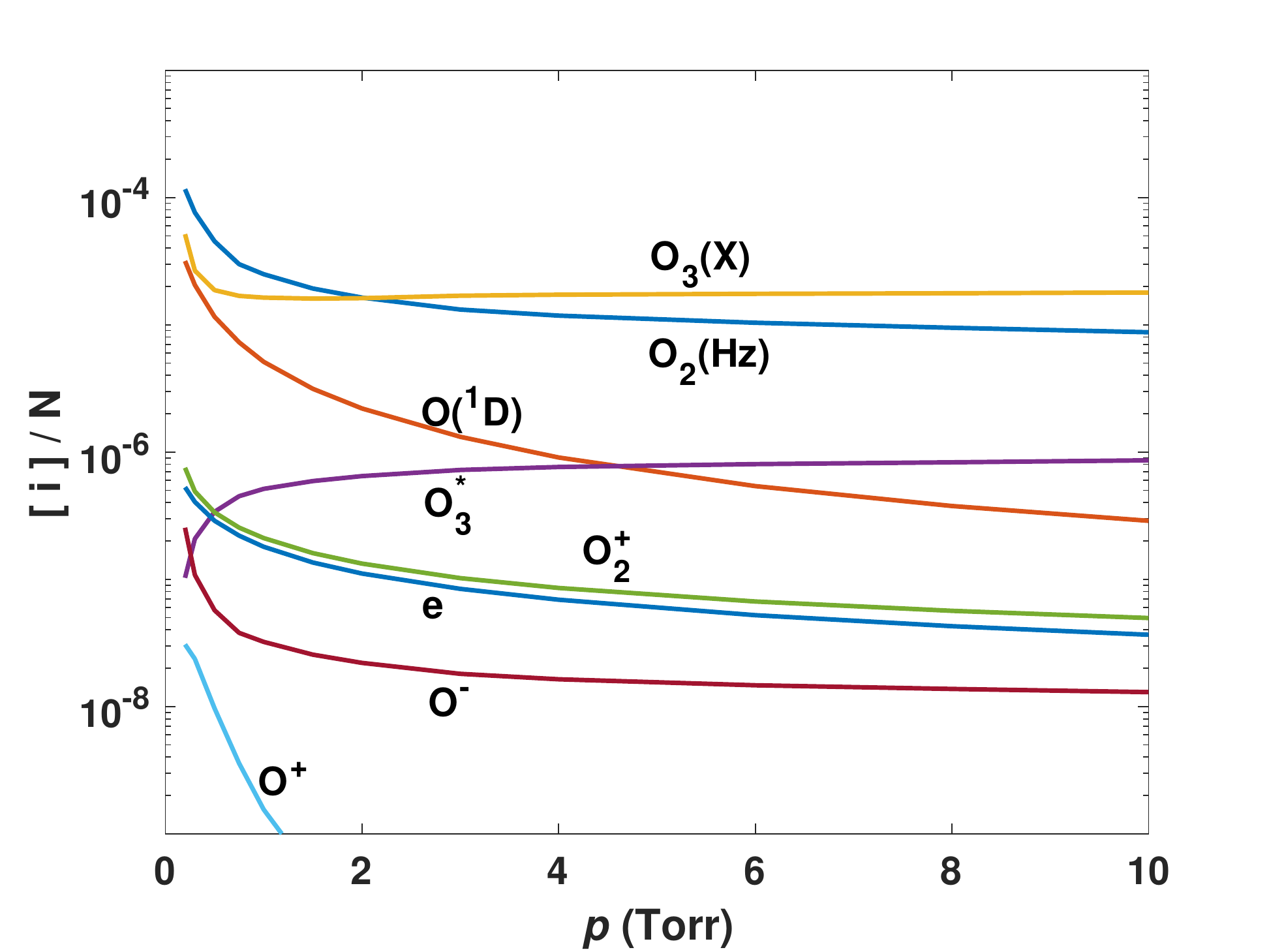}
    \subcaption{}
    \label{fig:MinorSpecies30mA}
  \end{minipage}
  \caption{Overview of the species fractions, $[i]/N$, in the discharge as a function of gas pressure, for a current of 30 mA.}
  \label{fig:species30mA}
\end{figure}

\quad\\
The fractions of species considered in the model (defined for species $i$ as $[i]/N$) are summarized in figure~\ref{fig:species30mA} as a function of gas pressure, for a discharge current of 30 mA. In this section we focus on the validation of the main species represented in figure~\ref{fig:MajorSpecies30mA}: $\mathrm{O_2(X)}$, $\mathrm{O_2(a)}$, $\mathrm{O_2(b)}$ and $\mathrm{O(^3P)}$.

\begin{figure}[H]
  \centering
  \begin{minipage}[H]{0.46\linewidth}
  \centering
     \includegraphics[width=0.95\linewidth]{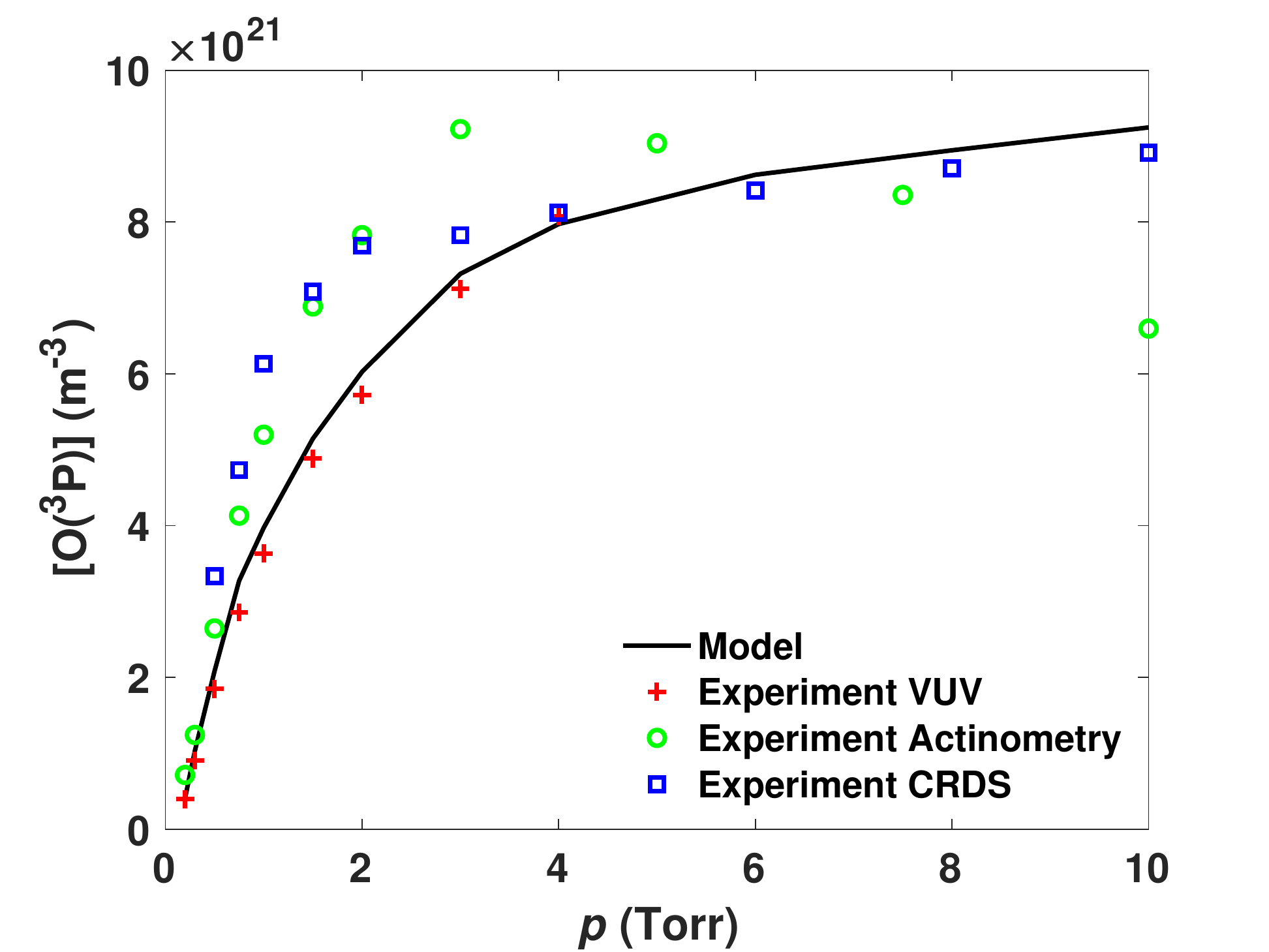}
    \subcaption{}
    \label{fig:O30mA}
  \end{minipage}
  \hfill
  \begin{minipage}[H]{0.46\linewidth}
  \centering
     \includegraphics[width=0.95\linewidth]{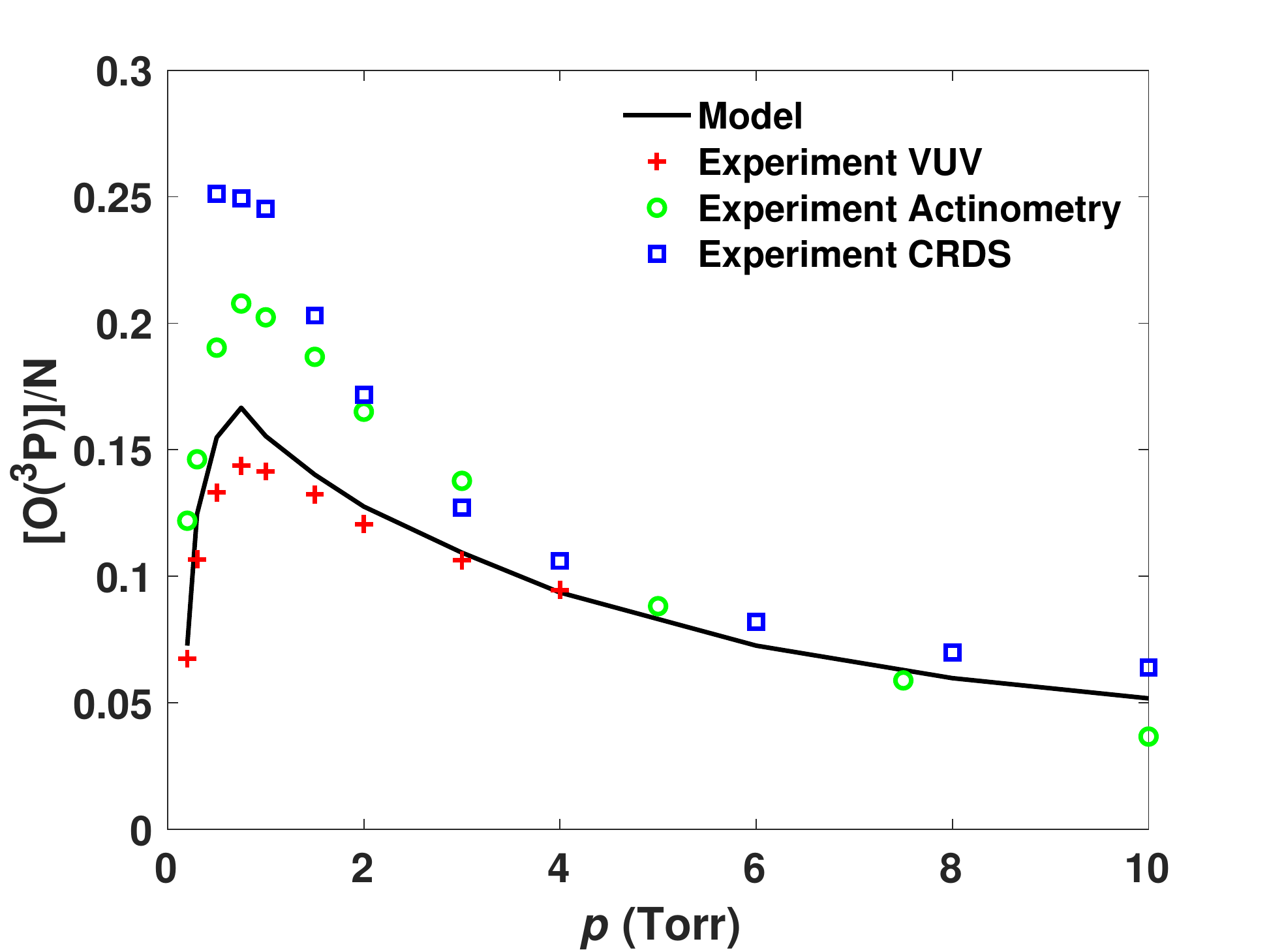}
    \subcaption{}
    \label{fig:OvsN30mA}
  \end{minipage}
  \caption{Comparison between modelling and experimental values of the $\mathrm{O(^3P)}$ (a) density and (b) fraction, as a function of gas pressure and for a current of 30 mA. The crosses, circles and squares represent measurements using VUV\cite{Booth_2020}, actinometry\cite{Booth_2019} and CRDS\cite{Booth_2022}, respectively.}
  \label{fig:OandOvsN30mA}
\end{figure}

The model calculations of the $\mathrm{O(^3P)}$ density and fraction are compared with experiment in figure~\ref{fig:OandOvsN30mA} for a discharge current of 30 mA. Take note that the vacuum ultraviolet (VUV)~\cite{Booth_2020} and actinometry~\cite{Booth_2019} measurements are radially averaged, whereas the cavity ringdown spectroscopy (CRDS)~\cite{Booth_2022} measurements are performed at the axis center. There is good accordance between the model and the experiment. Nevertheless, we should emphasize that the success in reproducing the experimental results is very much connected with the values used for the recombination probability $\gamma_\mathrm{O(^3P)}$, concerning reaction R73 of table~\ref{tab:heavyReactions}, taken from experiment~\cite{Booth_2020} and represented in figure~\ref{fig:gammaO_Booth2020}. For example, if we use the lower values of $\gamma_\mathrm{O(^3P)}$ from~\cite{Booth_2019} instead of~\cite{Booth_2020}, measured in similar conditions, the calculated $\mathrm{[O(^3P)]/N}$ has a similar shape but exhibits a maximum value at 1 Torr twice as large as shown here. This shows that the surface conditions in these discharges have poor reproducibility, and that further investigation is required from both modelling and experimental points of view to accurately address this phenomenon. The scattering between the oxygen density measurements obtained with actinometry~\cite{Booth_2019}, VUV~\cite{Booth_2020} and CRDS~\cite{Booth_2022} is another evidence for the low reproducibility of surface conditions and the inherent uncertainties in the experimental determination of the atomic oxygen concentration, more noticeable at low pressures. Furthermore, it makes clear that the validation of the model should be established more from the prediction of the correct dependences and orders of magnitude than by an exact agreement on the absolute values.

\begin{figure}[t]
	\centering
	\includegraphics[width=0.47\linewidth]{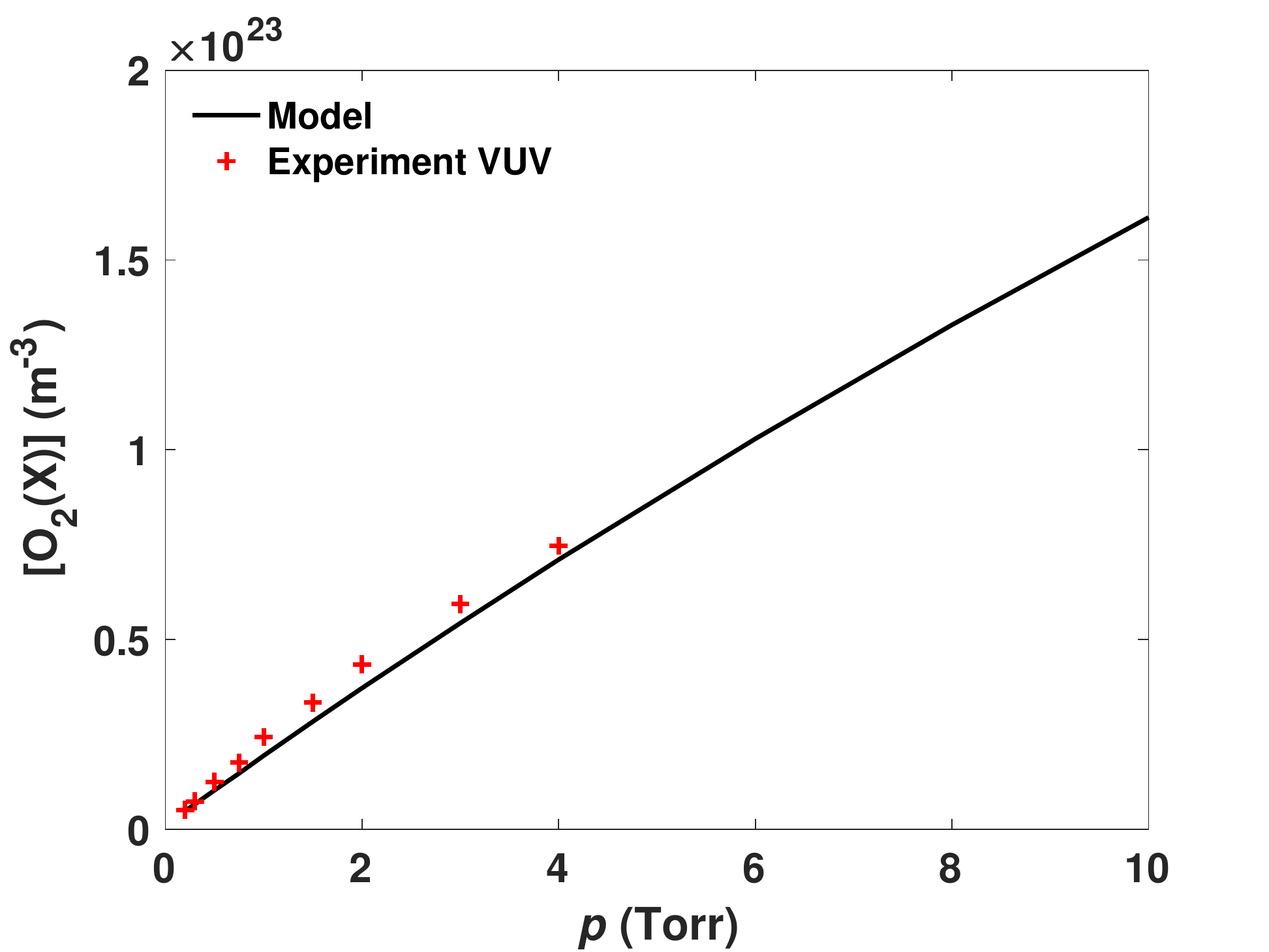}
	\caption{Comparison between modelling and VUV experimental values~\cite{Booth_2020} of the $\mathrm{O_2(X)}$ density.}
	\label{fig:O2X30mA}
\end{figure}

Figure~\ref{fig:O2X30mA} shows very good agreement between the modelling values and the VUV experimental measurements of $\mathrm{O_2(X)}$ density. Similarly to $\mathrm{O(^3P)}$, the model results are strongly correlated with $\gamma_\mathrm{O(^3P)}$, since $\mathrm{O_2(X)}$ and $\mathrm{O(^3P)}$ are the major species in the plasma~(cf. figure~\ref{fig:species30mA}).

\begin{figure}[t]
  \centering
  \begin{minipage}[H]{0.47\linewidth}
  \centering
     \includegraphics[width=\linewidth]{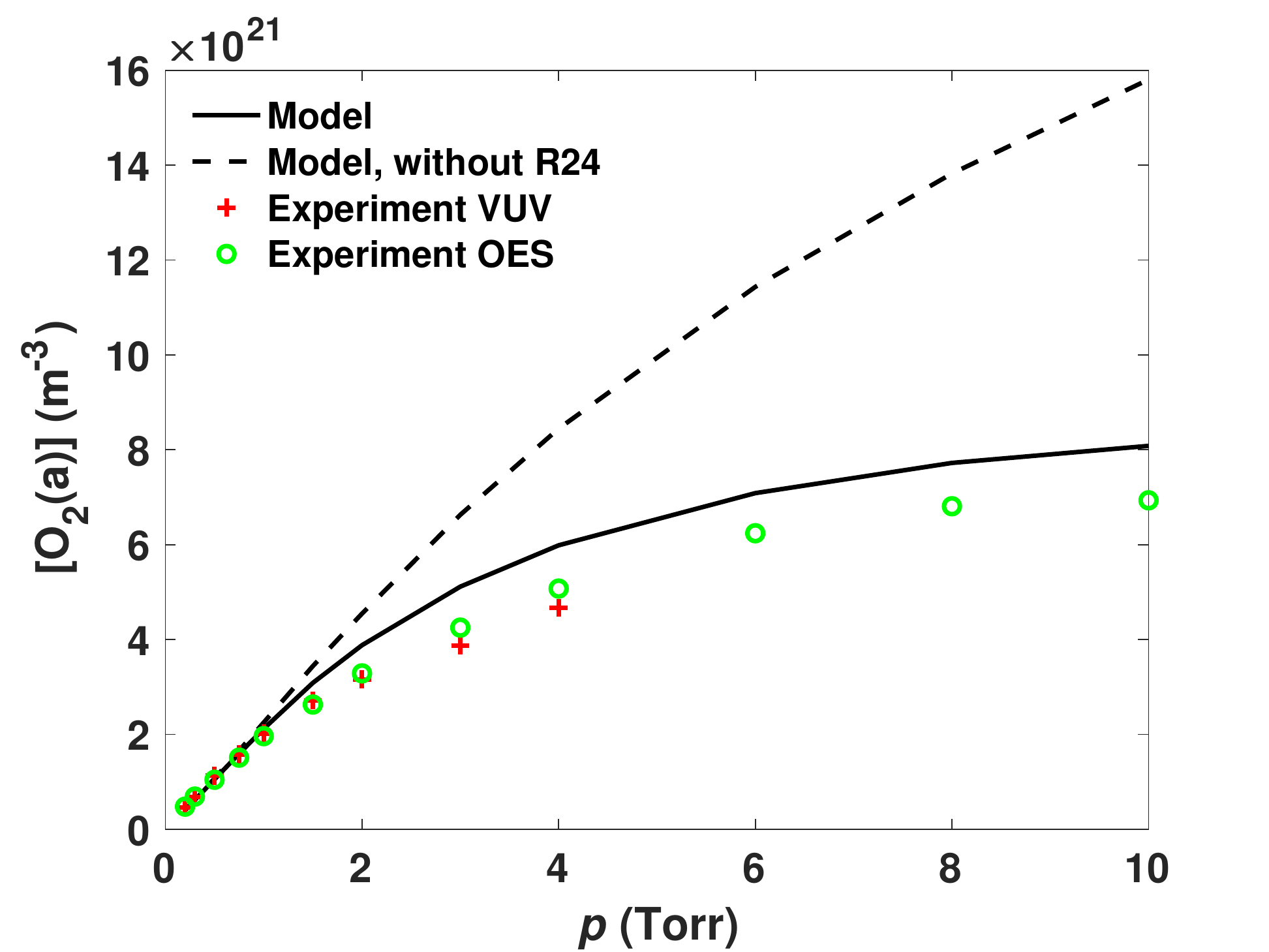}
    \subcaption{}
    \label{fig:O2a30mA}
  \end{minipage}
  \hfill
  \begin{minipage}[H]{0.47\linewidth}
  \centering
     \includegraphics[width=\linewidth]{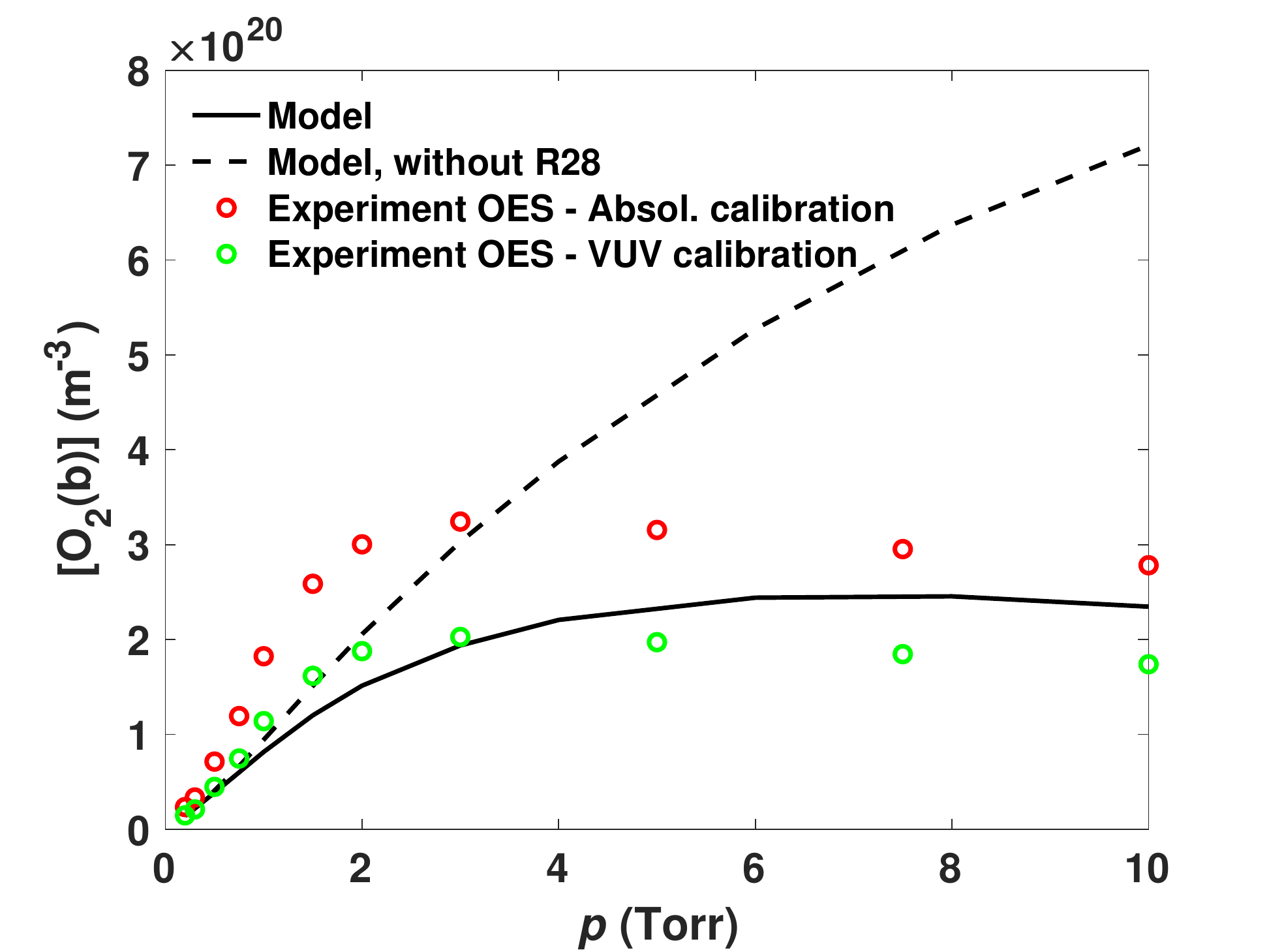}
    \subcaption{}
    \label{fig:O2b30mA}
  \end{minipage}
  \caption{Comparison between modelling and experimental\cite{Booth_2020,Booth_2022} values of the (a) $\mathrm{O_2(a)}$ and (b) $\mathrm{O_2(b)}$ densities, as a function of gas pressure, for a discharge current of 30 mA.}
  \label{fig:O2ab30mA}
\end{figure}

The model densities of $\mathrm{O_2(a)}$ and $\mathrm{O_2(b)}$ are compared with experiment in figure~\ref{fig:O2ab30mA}. For $\mathrm{O_2(a)}$, two sets of measurements from \cite{Booth_2020} are shown: VUV and optical emission spectroscopy (OES). The three-body quenching proposed by Braginskiy~\textit{et~al.}\cite{Braginskiy_2005} is fundamental to obtain a good agreement in trend and magnitude with experimental values, as can be seen by observing the results obtained when removing reaction R24. For $\mathrm{O_2(b)}$, one set of OES measurements is shown but with two different calibration/normalization factors~\cite{Booth_2022}: absolute OES calibration and scaling from VUV measurements. The deviation of 60\% between the two calibration factors gives an idea of the error associated with the absolute value of the measurement. Notice that, in this case, the reactive quenching proposed by Booth~\textit{et~al.}\cite{Booth_2022} is crucial to predict the density saturation after 3 Torr.

\begin{figure}[t]
  \centering
  \begin{minipage}[H]{0.47\linewidth}
  \centering
     \includegraphics[width=\linewidth]{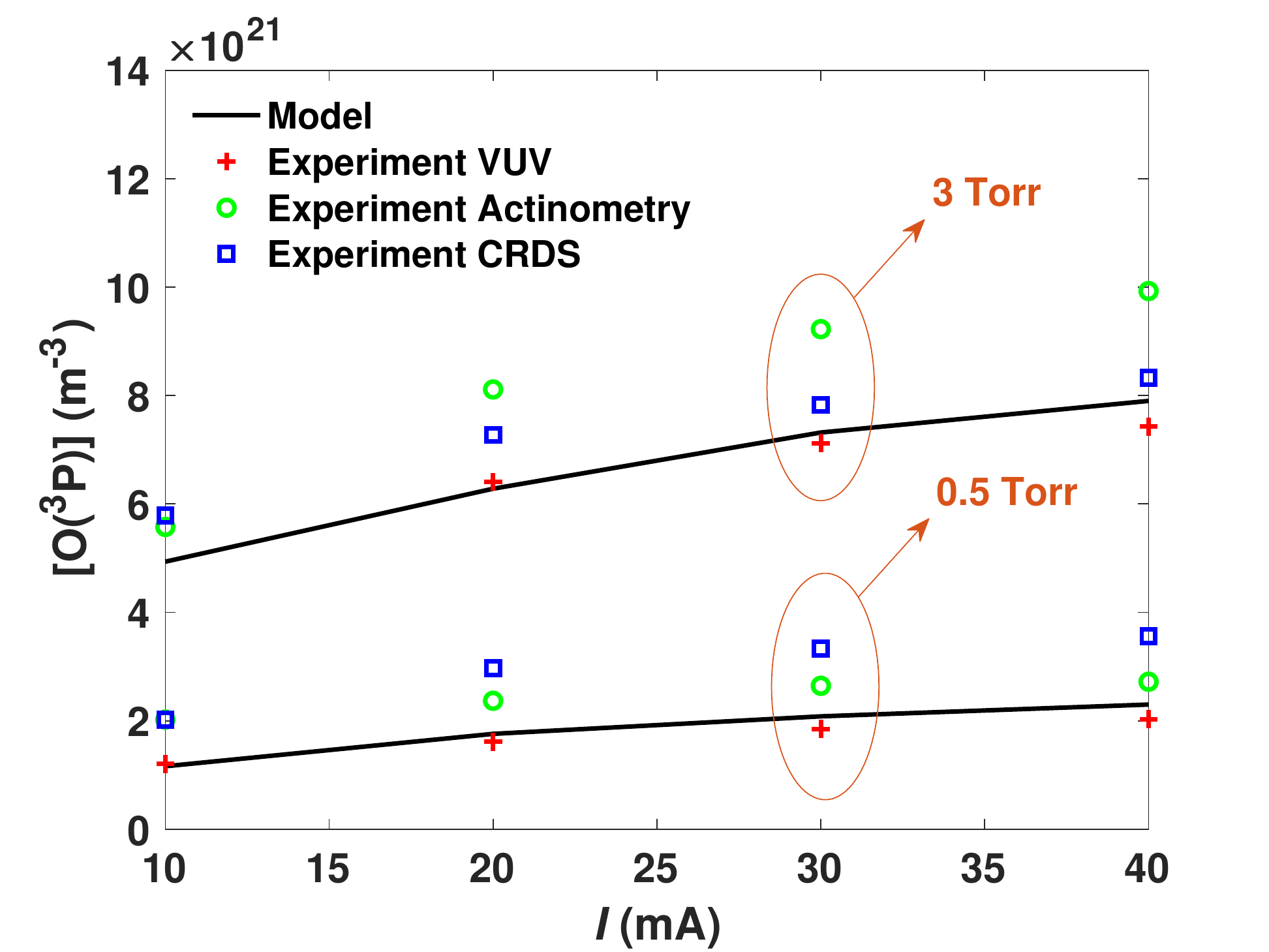}
    \subcaption{}
    \label{fig:mechO}
  \end{minipage}
  \hfill  
  \begin{minipage}[H]{0.47\linewidth}
  \centering
     \includegraphics[width=\linewidth]{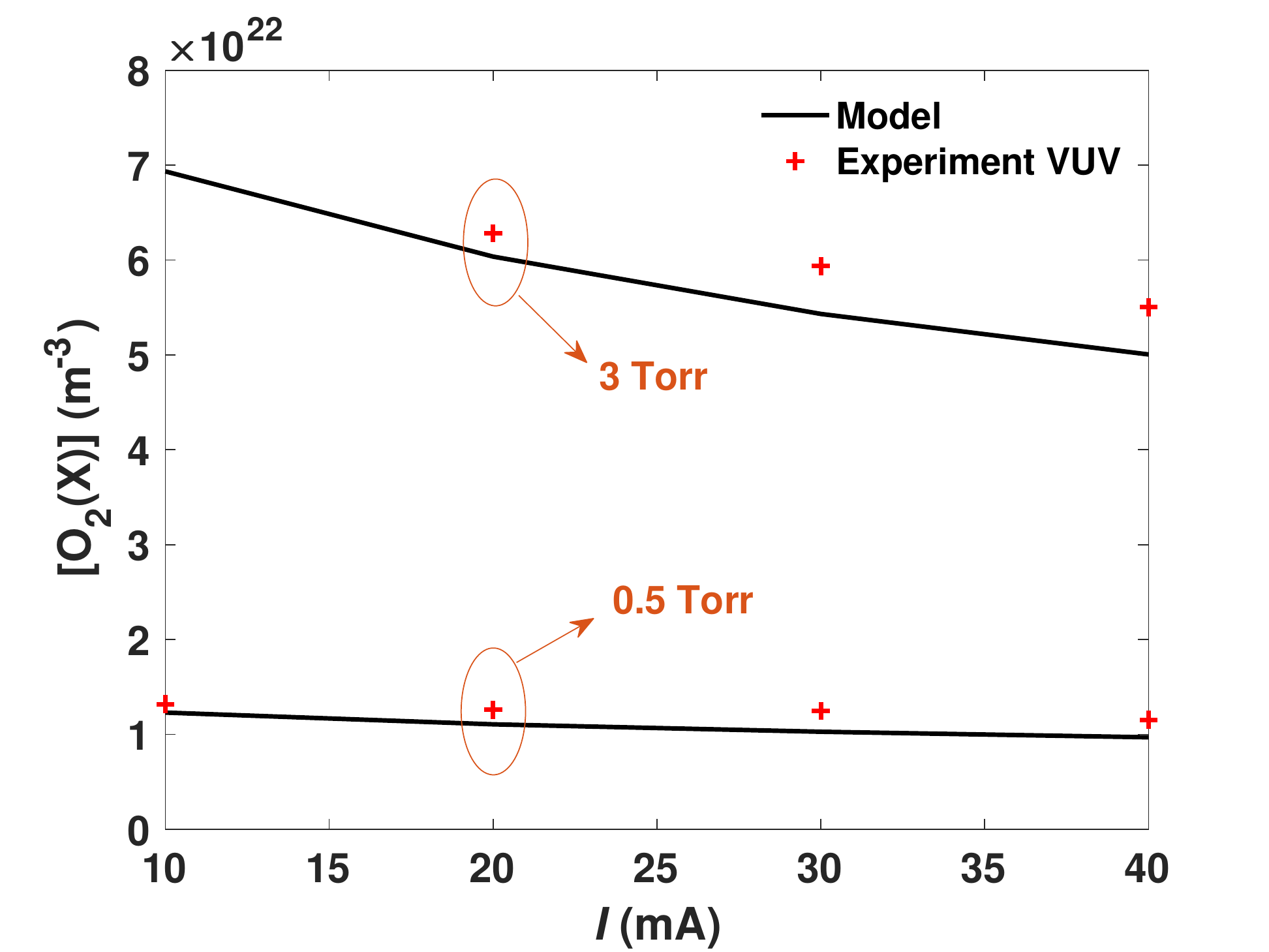}
    \subcaption{}
    \label{fig:mechO2X}
  \end{minipage}
  \begin{minipage}[H]{0.47\linewidth}
  \centering
     \includegraphics[width=\linewidth]{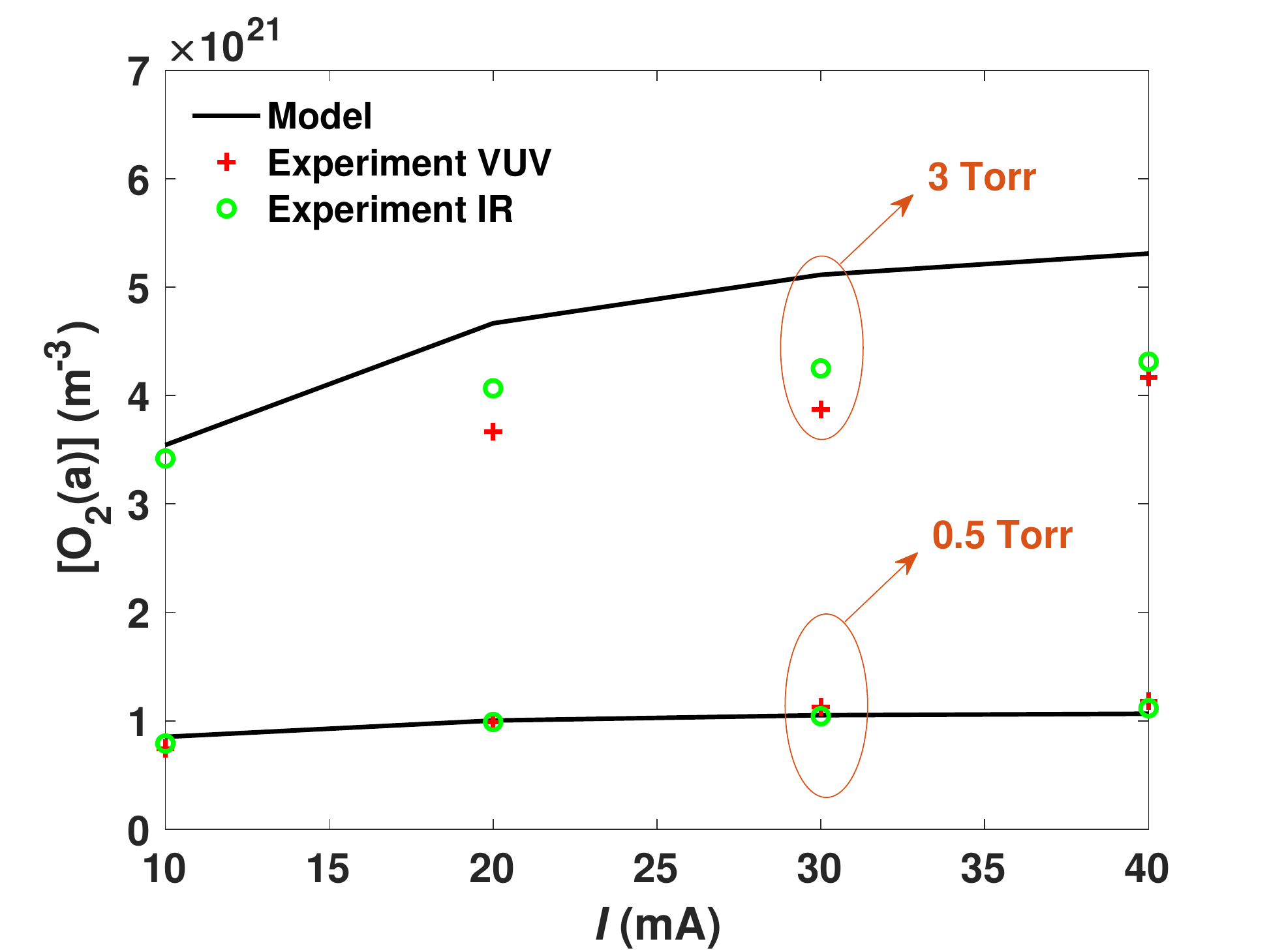}
    \subcaption{}
    \label{fig:mechO2a}
  \end{minipage}
  \hfill
  \begin{minipage}[H]{0.47\linewidth}
  \centering
     \includegraphics[width=\linewidth]{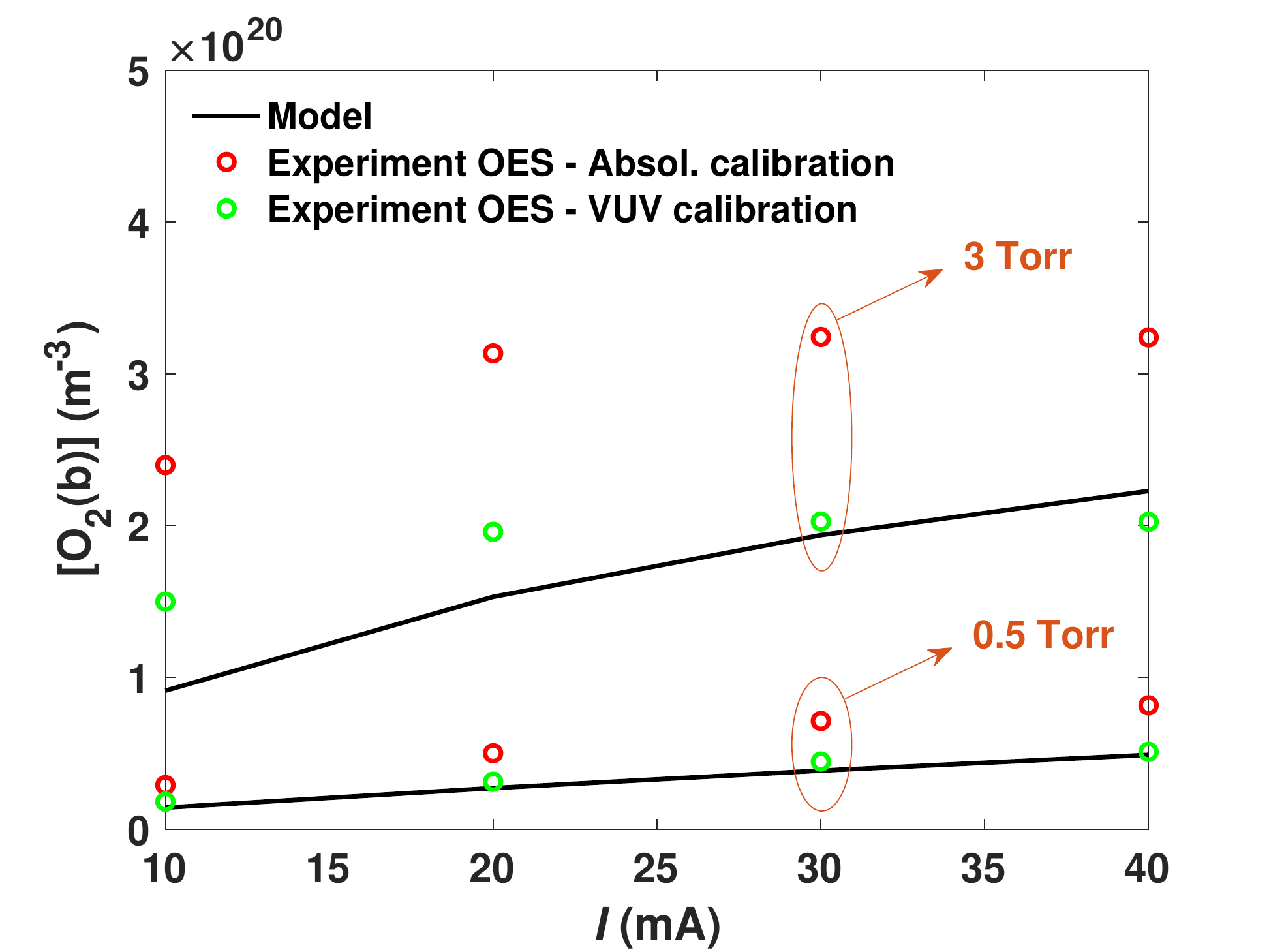}
    \subcaption{}
    \label{fig:mechO2b}
  \end{minipage}  
  \caption{Comparison between modelling and experimental values of the main species densities ($\mathrm{O(^3P)}$, $\mathrm{O_2(X)}$, $\mathrm{O_2(a)}$ and $\mathrm{O_2(b)}$), as a function of discharge current, for gas pressures of 0.5 and 3 Torr. Detailed references for the experimental measurements~\cite{Booth_2019,Booth_2020,Booth_2022} are mentioned in figures \ref{fig:OandOvsN30mA}-\ref{fig:O2ab30mA} and the text.}
  \label{fig:speciesVarCurrent}
\end{figure}

The effect of discharge current on the main species densities is summarized in figure~\ref{fig:speciesVarCurrent}, for gas pressures of 0.5 and 3 Torr. As expected, an increase in current / electron density leads to a larger dissociation - more $\mathrm{[O(^3P)}]$ and less $\mathrm{[O_2(X)]}$ - and to higher densities of the metastables $\mathrm{O_2(a)}$ and $\mathrm{O_2(b)}$. Besides, the model captures very well the tendencies and the absolute values found in experiment.

\pagebreak
\subsection{Main processes}\label{sec:mainProcesses}

Benefiting from the reaction mechanism of oxygen discharges validated in section \ref{sec:validation}, we can identify the main processes controlling the kinetic behavior of the system. Here, we highlight the most important reactions for the production/destruction of the main species and the dominant sources for gas heating.

\subsubsection{Species production/destruction}\label{sec:mainSpeciesProcesses}
\begin{figure}[H]
  \centering
  \begin{minipage}[H]{0.49\textwidth}
  \centering
     \includegraphics[width=1.04\textwidth]{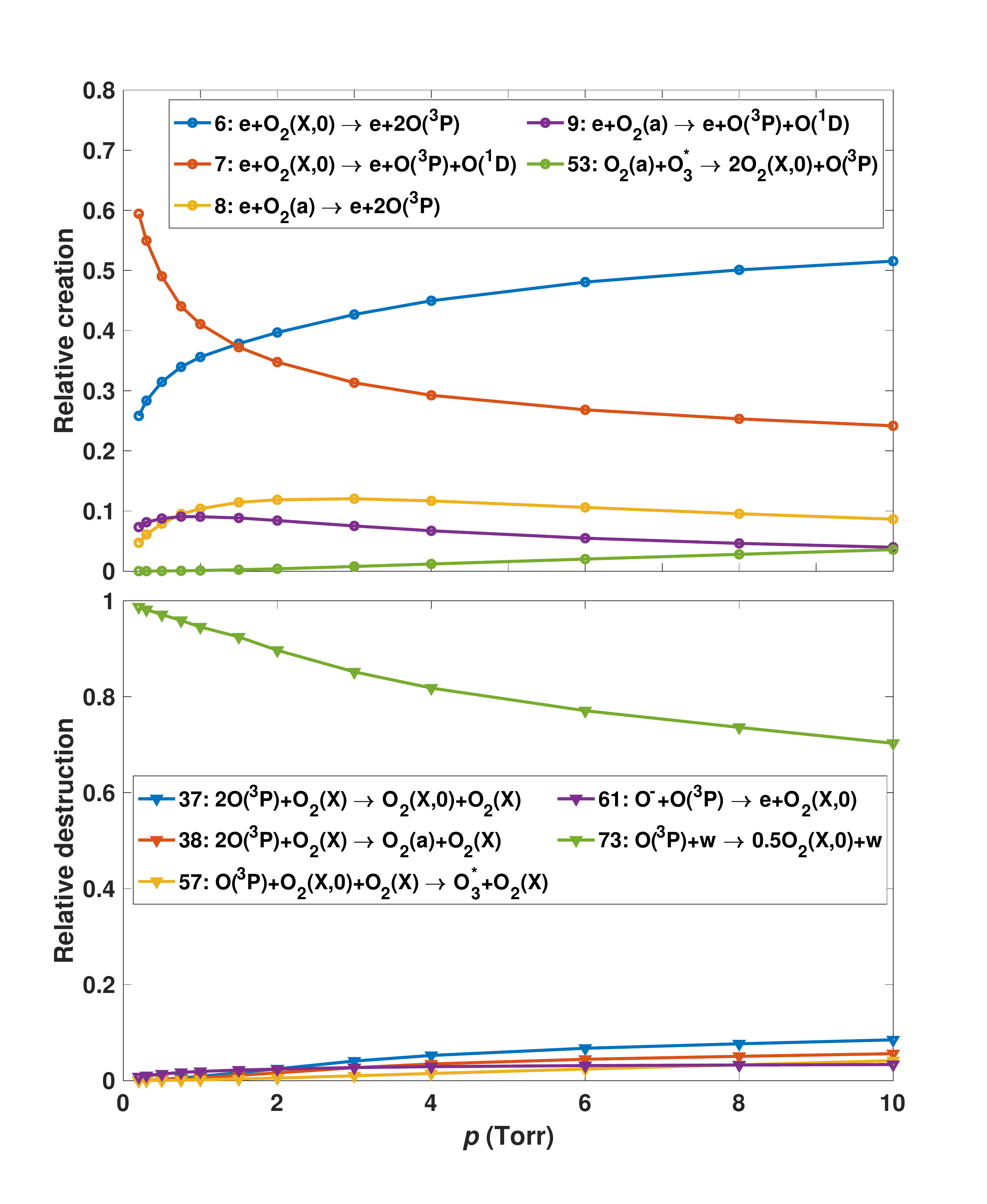}
    \subcaption{}
    \label{fig:mechO}
  \end{minipage}
  \hfill  
  \begin{minipage}[H]{0.49\textwidth}
  \centering
     \includegraphics[width=1.04\textwidth]{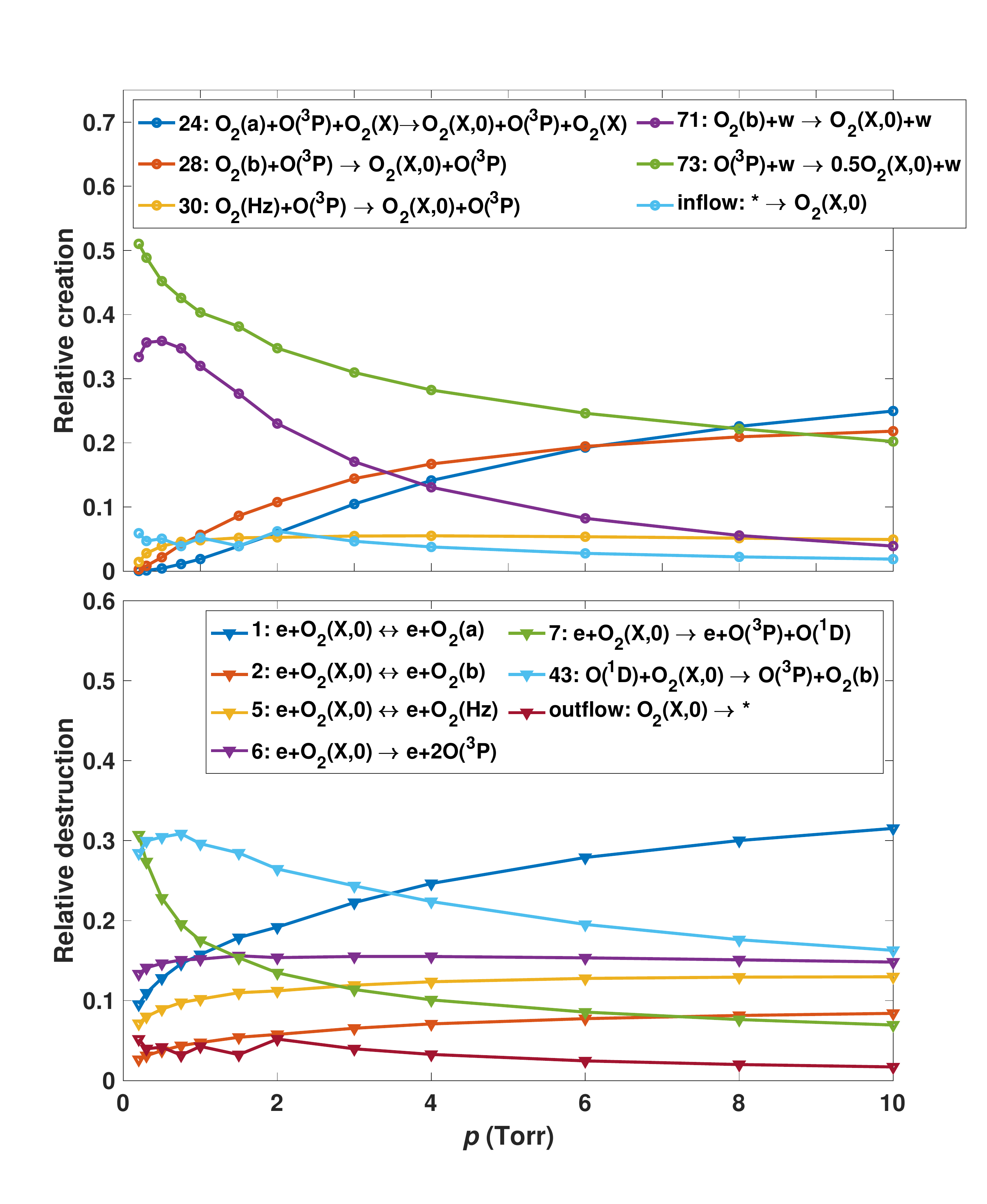}
    \subcaption{}
    \label{fig:mechO2X}
  \end{minipage}
  \\
  \begin{minipage}[H]{0.49\textwidth}
  \centering
     \includegraphics[width=1.04\textwidth]{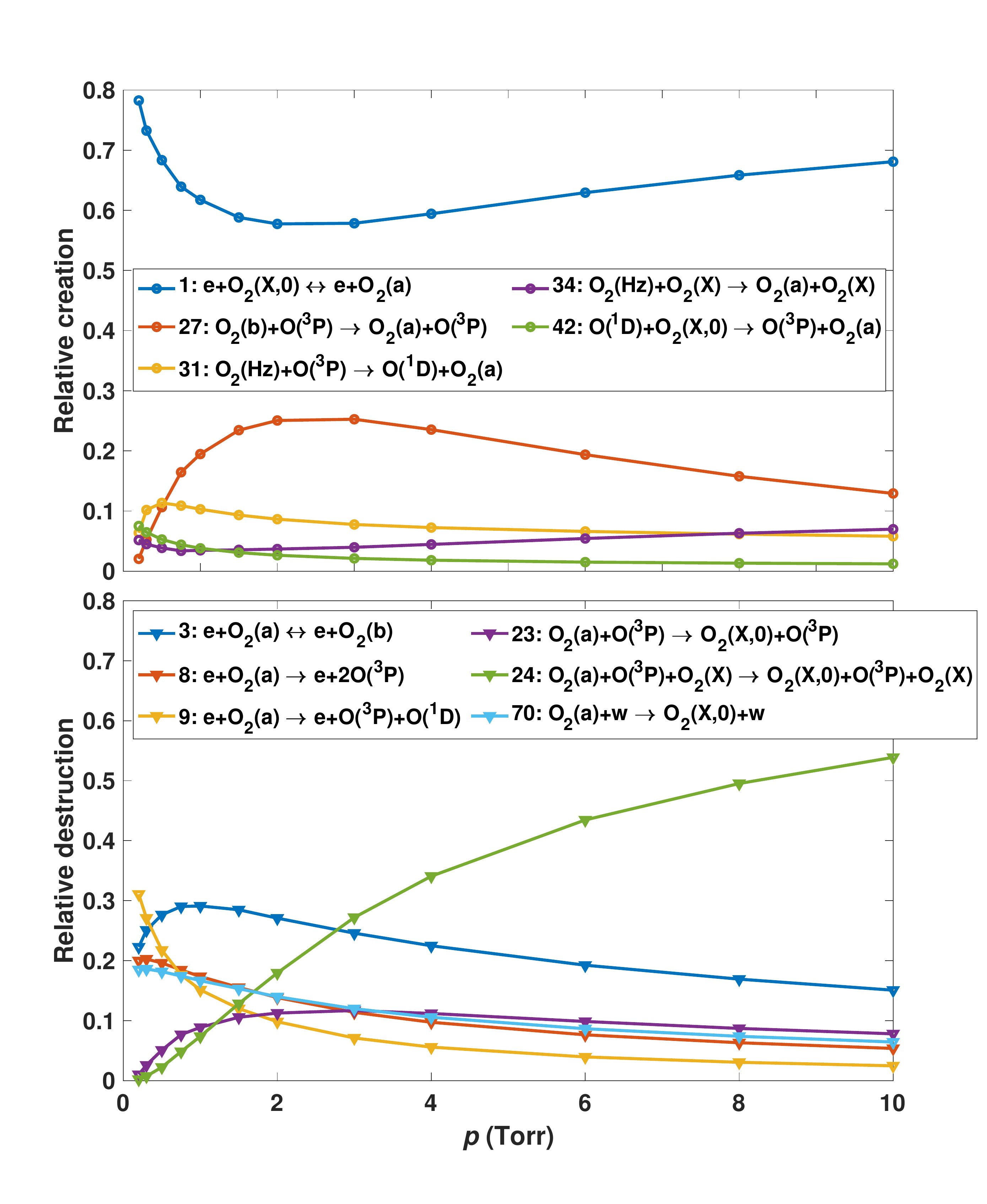}
    \subcaption{}
    \label{fig:mechO2a}
  \end{minipage}
  \hfill
  \begin{minipage}[H]{0.49\textwidth}
  \centering
     \includegraphics[width=1.04\textwidth]{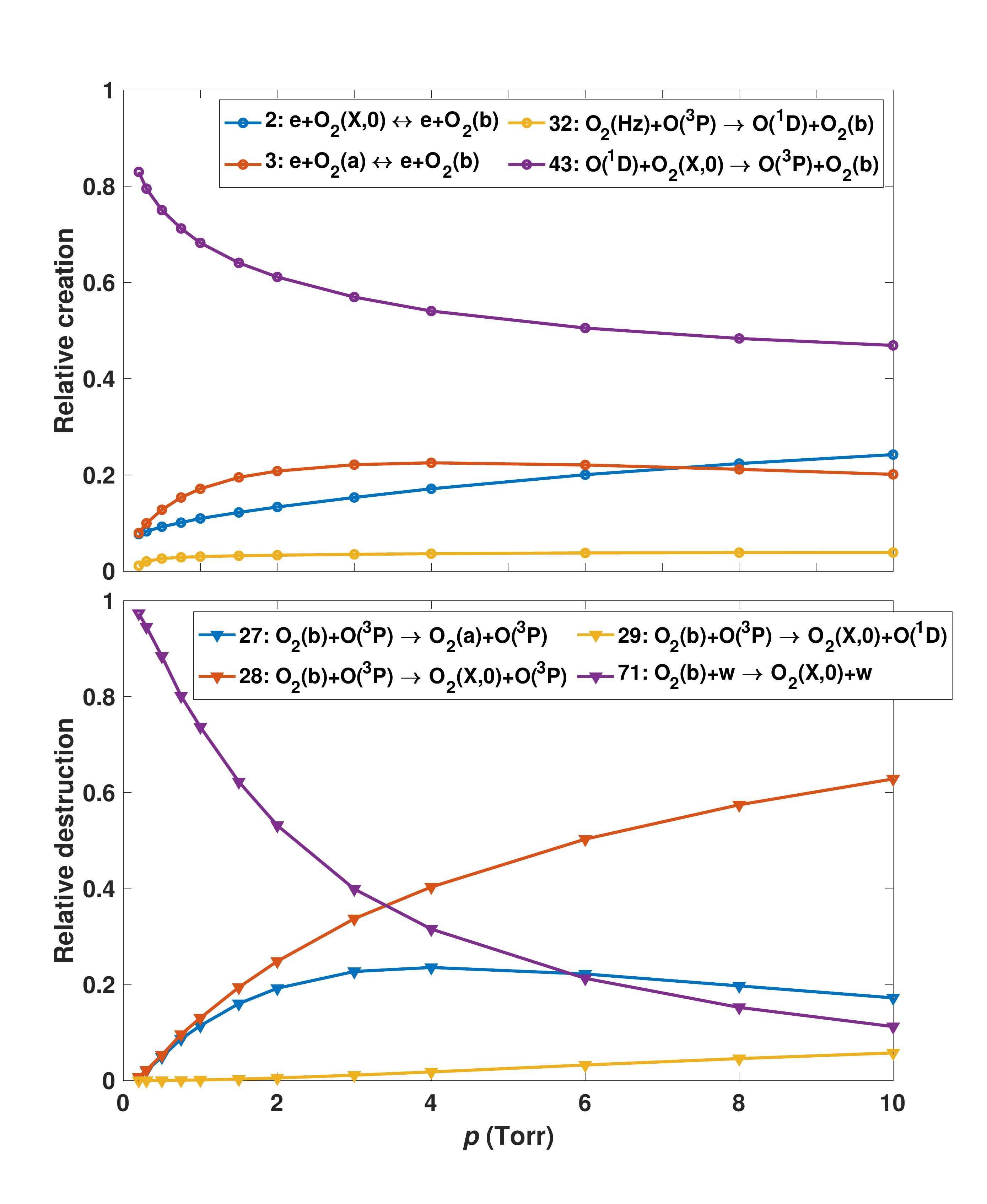}
    \subcaption{}
    \label{fig:mechO2b}
  \end{minipage}
  \caption{Main processes for the creation (full circles) and  destruction (full triangles) of (a) atomic oxygen, (b) $\mathrm{O_2(X)}$, (c) $\mathrm{O_2(a)}$ and (d) $\mathrm{O_2(b)}$, as a function of pressure, for a current of 30 mA. In the legends, ``wall'' is abbreviated as ``w''.}
  \label{fig:mechAll30mA}
\end{figure}

\quad\\
The main processes for the production/destruction of neutral atomic oxygen, $\mathrm{O_2(X)}$, $\mathrm{O_2(a)}$ and $\mathrm{O_2(b)}$ are represented in figure~\ref{fig:mechAll30mA}. The species $\mathrm{O(^3P)}$ and $\mathrm{O(^1D)}$ are considered together in this analysis, since there are strong excitation and quenching processes between them that do not lead to an effective destruction or production of O atoms, merely redistributing the population of these two states~\cite{Annusova_2018}. Similarly, the analysis for $\mathrm{O_2(X)}$ joins the processes due to all vibrational states, eliminating reactions that only lead to vibrational exchanges. In processes with double arrow ($\leftrightarrow$), the contribution reflects the net value of the direct and reverse components. On a general note, we see that the electron-impact processes and the wall de-excitation/recombination play a key role in the kinetics of all species.

In figure~\ref{fig:mechO}, we observe that the atomic oxygen is mostly created by electron impact with $\mathrm{O_2(X)}$ and $\mathrm{O_2(a)}$ (R6-R9), and it is mainly destroyed by $\mathrm{O(^3P)}$ recombination at the wall (R73). For higher pressures, three-body recombination processes start to be important, having an impact of $\sim20\%$ on the $\mathrm{O}$ atom destruction at 10 Torr.

Figure~\ref{fig:mechO2X} evidences that $\mathrm{O_2(X)}$ production is mostly controlled by $\mathrm{O(^3P)}$ recombination at the wall (R73), quenchings of $\mathrm{O_2(a)}$ (R24), $\mathrm{O_2(b)}$ (R28) and $\mathrm{O_2(Hz)}$ (R30), and de-excitation of $\mathrm{O_2(b)}$ at the wall (R71). The main processes for $\mathrm{O_2(X)}$ destruction are electron-impact excitations (R1, R2 and R5), electron-impact dissociations (R6 and R7) and $\mathrm{O(^1D)}$ quenching (R43). Note that, for the conditions explored in this work, the inflow and outflow processes do not play a major role, having an influence in the production/destruction of $\mathrm{O_2(X)}$ of less than 8\% for all conditions.

As seen in figure~\ref{fig:mechO2a}, the main production of $\mathrm{O_2(a)}$ is given by electron-impact excitation from the ground state (R1) and quenching of $\mathrm{O_2(b)}$ with $\mathrm{O(^3P)}$ (R27). The quenching of $\mathrm{O_2(Hz)}$ through  R31 and R34 is also relevant, with a summed contribution of up to 15\%. The $\mathrm{O_2(a)}$ destruction is controlled by electron-impact processes (R3, R8 and R9), two- and three- body quenchings (R23 and R24), and de-excitation at the wall (R70). Reaction R24 is fundamental at high pressures, as shown also in figure \ref{fig:O2a30mA}, contributing more than 50\% for the destruction at 10 Torr.

Finally, in figure~\ref{fig:mechO2b}, we find that the $\mathrm{O_2(b)}$ creation is mostly associated with electron-impact excitations (R2 and R3) and specially with the quenching of $\mathrm{O(^1D)}$ (R43). For low pressures, the destruction occurs predominantly through wall deactivation (R71), and for higher pressures, through quenching with $\mathrm{O(^3P)}$~(R27~-~R29).

\begin{figure}[H]
	\centering
	\includegraphics[width=0.47\linewidth]{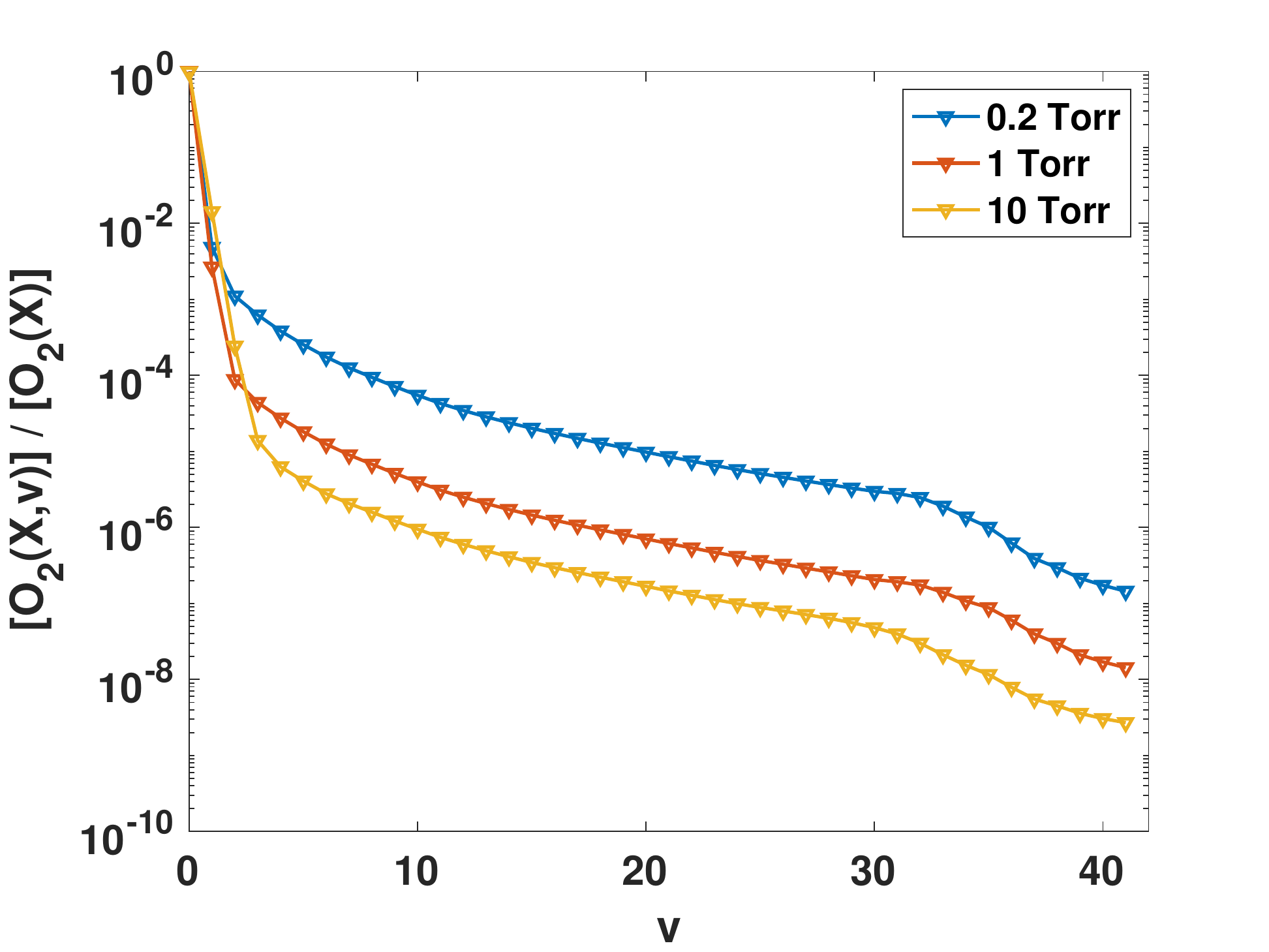}
	\caption{VDF of $\mathrm{O_2(X,v)}$ for different pressures and a current of 30 mA.}
	\label{fig:VDFs}
\end{figure}

None of the important reactions presented in figure \ref{fig:mechAll30mA} involve vibrationally-excited states of $\mathrm{O_2(X)}$, although they are considered in the calculations. This may be understood by looking at the vibrational distribution function (VDF) of $\mathrm{O_2(X,v})$, represented in figure~\ref{fig:VDFs} for different pressures and a current of 30 mA. The VDFs are in rather good agreement with values measured under somewhat similar conditions, see figure 9 of \cite{Guerra_2019}. For all conditions, the relative populations of $\mathrm{v=1}$ and 2 are around $10^{-2}$ and $10^{-3}$, respectively. Therefore, the influence of vibrationally-excited levels on the chemistry is very limited, at least under the conditions studied in this work. However, the inclusion of vibrational kinetics is fundamental for an accurate self-consistent calculation of the gas temperature due to its relevance in the gas heating, as shown in the next section.

\begin{figure}[H]
	\centering
	\includegraphics[width=0.9\linewidth]{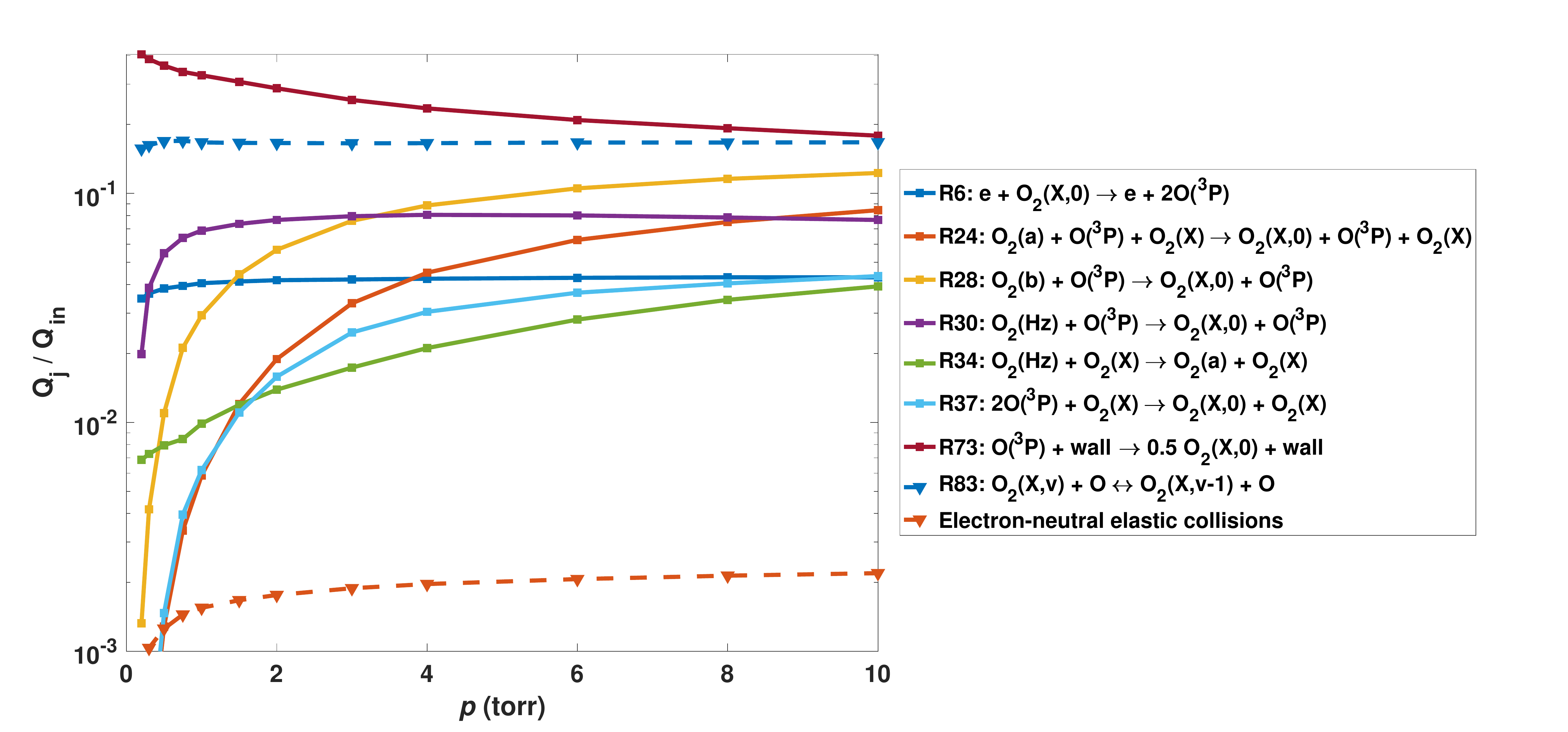}
  \caption{Relative contributions, $Q_j/Q_\mathrm{in}$, of the most relevant kinetic processes for gas heating, as a function of pressure for a current of 30 mA.}
	\label{fig:gasHeatingMech}
\end{figure}

\subsubsection{Gas heating}\label{sec:mainHeatProcesses}\quad\\
The relative contributions, $Q_j/Q_\mathrm{in}$ (cf. equation \ref{eq:heatTransferAverage}), of the most relevant kinetic processes for gas heating, are shown in figure~\ref{fig:gasHeatingMech}. As discussed in section~\ref{sec:LoKIC}, each source term $Q_j$ is calculated by multiplying the corresponding reaction rate by its enthalpy ($\epsilon_\mathrm{gh}$). Although the gas heating due to electron-neutral elastic collisions is accounted for the calculation of $Q_\mathrm{in}$, its relative contribution is always below $3\times10^{-3}$. In the entire pressure range, the dominant heating process is given by $\mathrm{O(^3P)}$ recombination on the wall. Recall that for processes occurring at the wall, we assume that half of the energy released goes to the plasma volume ($f_\mathrm{wall}=0.5$), the rest being dissipated to the wall (cf. section~\ref{sec:LoKIC}). Therefore, this assumption  has a significant influence on the results. For example, for a current of 30 mA and a gas pressure of 1 Torr, the gas temperature is 358, 378 and 396 K, for values of $f_\mathrm{wall}= 0$, 0.5 and 1, respectively; for 30 mA and 10 Torr, $T_\mathrm{g}$ is 514, 540 and 565 K, for the same values of $f_\mathrm{wall}$.

The V-T transfers by collisions with O atoms (R83) also play a vital role in gas heating, with a relative contribution of $\sim 17\%$. For this reason, when the gas temperature is calculated self-consistently in an $\mathrm{O_2}$ discharge model, the kinetics of vibrationally-excited states $\mathrm{O_2(X,v)}$ should be properly accounted.

Finally, the quenchings of $\mathrm{O_2(a)}$ (R24), $\mathrm{O_2(b)}$ (R28) and $\mathrm{O_2(Hz)}$ (R30 and R34) also make a very relevant contribution to heating, which grows with pressure, and the three-body recombination of $\mathrm{O(^3P)}$ (R37) starts to have a small influence at high pressure.

\pagebreak
\section{Conclusions}\label{sec:conclusions}

In this work, we present a \textit{reaction mechanism} for oxygen plasmas. The reaction mechanism was validated by comparing available experimental values~\cite{Booth_2019,Booth_2020,Booth_2022} against modelling calculations of the reduced electric field, the gas temperature and the densities of the predominant species in the glow discharge - $\mathrm{O_2(X)}$, $\mathrm{O_2(a)}$, $\mathrm{O_2(b)}$ and $\mathrm{O(^3P)}$ - at gas pressures between 0.2 and 10 Torr. Moreover, we have verified that the present oxygen reaction mechanism still describes correctly our previous works concerning the kinetics of ground-state and vibrationally-excited ozone~\cite{Marinov_2013} and vibrationally-excited $\mathrm{O_2(X,v)}$~\cite{Annusova_2018}.

A detailed study of the main processes for species creation/destruction quantified the relative importance of electron-impact excitation/dissociation, wall de-excitation/recombination and two- and three- body quenching processes. The three-body quenching (R24, $\mathrm{O_2(a) + O(^3P) + O_2(X) \rightarrow O_2(X,0) + O(^3P) + O_2(X)}$) proposed by Braginskiy \textit{et al.}\cite{Braginskiy_2005} is instrumental to reproduce the correct behavior of $\mathrm{O_2(a)}$ at high pressure, while electron-impact processes ($\mathrm{e+O_2(X)\leftrightarrow e+ O_2(a) \leftrightarrow e + O_2(b)}$) are prevalent at low-pressures. The non-reactive and reactive quenching of $\mathrm{O_2(b)}$ (R28, $\mathrm{O_2(b) + O(^3P) \rightarrow O_2(X,0) + O(^3P)}$) ~\cite{Booth_2022} allows obtaining the saturation of its density with increasing pressure; deactivation at the wall is the main destruction process at low pressure; and reaction R43 involving $\mathrm{O(^1D)}$ constitutes always an important creation mechanism. The kinetics of $\mathrm{O(^3P)}$ is strongly controlled by wall recombination, and its density is highly dependent on the variation of the recombination probability with the wall temperature, gas pressure and discharge current. %Besides, through an analysis of the main mechanisms responsible for gas heating, we found that the $\mathrm{O(^3P)}$ recombination at the wall, the V-T transfers with O atoms and the quenching of electronically-excited states are contributing the most for the rising of gas temperature.

The gas temperature plays a key role in the modelling, since many rate coefficients are temperature-dependent. The calculation of the gas temperature can be done with a self-consistent thermal model accounting for the energy released in volume and wall reactions, the heat losses due to thermal conduction and the gas-to-wall convection transfer. It can also adopt alternative (simpler) approaches to introduce a temperature jump from the gas to wall. In any case, $T_\mathrm{g}$ should be experimentally monitored and model predictions for $T_\mathrm{g}$ should agree satisfactorily with the measurements. In the present conditions, the $\mathrm{O(^3P})$ wall recombination, the V-T transfers to O atoms and the quenching of $\mathrm{O_2}$ electronic states, including the Herzberg states $\mathrm{O_2(A^{\prime 3}\Delta_u,A^3\Sigma_u^+,c^1\Sigma_u^-)}$, are the main contributors for the gas heating. 

We believe that the low-temperature plasma community should make an effort to adopt a paradigm in the development of kinetic schemes to study plasma chemistry, assuming the same rigour and systematization in validation devoted to the development of complete and consistent electron-impact cross section sets, and embracing the good practices from other communities. In particular, each kinetic scheme should be compared against dedicated benchmark experiments. In addition, an experimental effort to generate and organize such experimental data would be welcome. In the case of oxygen discharges, the systematic measurements from LPP and MSU~\cite{Booth_2019,Booth_2020,Booth_2022} constitute a very complete experimental data set and can be used as a validation test for other models. However, more experimental data from various independent researchers is necessary to evaluate possible problems of reproducibility or systematic errors and to extend the conditions for validation. 

It should be noted that the validation of a kinetic scheme by trying to put all modelling results on top of the experimental values risks to be meaningless. On the one hand, experimental measurements have uncertainties associated, which in some cases are difficult to quantify. On the other hand, model equations comprise various approximations. For example, in 0D models, the assumption of certain density profiles to describe the transport of neutral and charged species allows to obtain the correct trends and identify the main elementary processes ruling the discharge, but sometimes the absolute values are somewhat inaccurate~\cite{Viegas_2023}. For these reasons, it is more relevant to ensure that the kinetic scheme is physically consistent and describes the observed tendencies of several quantities in a wide range of conditions.

Future work will include analysis on different pressure ranges and other discharge configurations. By extending the domain of validity of the reaction mechanism, as distinct dominant elementary processes may be at play with rate coefficients for different temperature ranges, the modelling results will necessarily become more predictive and will involve the use of fewer tuning parameters. Moreover, a surface kinetics model will soon be coupled to the volume kinetics, in order to obtain an extended self-consistent model that does not rely on experimental data for the recombination probabilities. This will allow for a deeper understanding of the experimental results and which physical parameters are controlling the surface reactions. Finally, the impact of the two-term approximation used in LoKI-B on the plasma chemistry results will be quantified, by comparison with the results obtained when using the electron kinetics Monte Carlo solver LoKI-MC~\cite{Dias_2023}, in oxygen and other relevant gases. % Sensitivity analysis tools~\cite{Terraz_2020} may be instrumental in bringing further insight and fine tuning reaction mechanisms.

\section*{Acknowledgments}
\noindent
This work was partially supported by the Portuguese FCT, under grant PD/BD/150414/2019 (PD-F APPLAuSE) and Projects UIDB/50010/2020, UIDP/50010/2020, EXPL/FIS-PLA/0076/2021 and MIT-EXPL/ACC/0031/2021, by the European Union's Horizon 2020 research and innovation programme, under grant agreement MSCA ITN 813393, and by the European Union (NextGenerationEU) and the Spanish Ministerio de Universidades (Plan de Recuperacion, Transformación y Resiliencia).\\
We would like to thank Carlos D Pintassilgo for fruitful discussions on the thermal model. We are also deeply thankful to Jean-Paul Booth, Olivier Guaitella, Dmitry Voloshin, Dmitry Lopaev and Tatyana Rakhimova for invaluable discussions on the heavy-species processes and for providing the experimental data presented in figures \ref{fig:EN} and \ref{fig:TgasAver}. Finally, we acknowledge the careful revision and suggestions by the referees, which helped us to improve the manuscript.

\section*{References}
\bibliography{O2DCPaper.bib}

\end{document}

%% file: oxygenElecImpactScheme.in
\begin{table}[]
\centering
\scriptsize
\caption{List of electron-impact processes considered in this work for vibrationally-cold oxygen. New processes compared to \cite{Annusova_2018} are marked with **.  Processes with rate coefficients updated in this work, with respect to \cite{Annusova_2018},  are marked with *. $\epsilon_\mathrm{gh}$ is the energy released for gas heating.}
\label{tab:eImpactReactions}
\begin{tabular}{@{}lllll@{}}
\toprule
Reaction &
   &
  Rate coefficient ($\mathrm{m^3   s^{-1}}$) &
  $\epsilon_\mathrm{gh}$ (eV) &
  Ref. \\ \midrule
\multicolumn{2}{l}{\textit{Electron-impact   excitation}} &
   &
   &
   \\
R1 &
  $\mathrm{e + O_2(X,0) \leftrightarrow e + O_2(a)}$ &
  $f(E/N)$ &
  $-$ &
  \cite{Phelps_1985_crossSections,Oxyg_ISTLisbon} \\
R2 &
  $\mathrm{e + O_2(X,0) \leftrightarrow e + O_2(b)}$ &
  $f(E/N)$ &
  $-$ &
  \cite{Phelps_1985_crossSections,Oxyg_ISTLisbon} \\
R3 &
  $\mathrm{e + O_2(a) \leftrightarrow e + O_2(b)}$ &
  $f(E/N)$ &
  $-$ &
  \cite{Tashiro_2006,Oxyg_ISTLisbon} \\
R4 &
  $\mathrm{e + O(^3P) \leftrightarrow e + O(^1D)}$ &
  $f(E/N)$ &
  $-$ &
  \cite{Laher_1990,Oxyg_ISTLisbon} \\
R5$^{**}$ &
  $\mathrm{e + O_2(X,0) \leftrightarrow e + O_2(Hz)}$ &
  $f(E/N)$ &
  $-$ &
  \cite{Phelps_1985_crossSections,Oxyg_ISTLisbon} \\
\multicolumn{2}{l}{\textit{Electron-impact   dissociation}} &
   &
   &
   \\
R6 &
  $\mathrm{e + O_2(X,0) \rightarrow e + 2O(^3P)}$ &
  $f(E/N)$ &
  0.84 &
  \cite{Phelps_1985_crossSections,Oxyg_ISTLisbon} \\
R7 &
  $\mathrm{e + O_2(X,0) \rightarrow e + O(^3P) + O(^1D)}$ &
  $f(E/N)$ &
  1.27 &
  \cite{Phelps_1985_crossSections,Oxyg_ISTLisbon} \\
R8 &
  $\mathrm{e + O_2(a) \rightarrow e + 2O(^3P)}$ &
  $f(E/N)$ &
  0.84 &
  \cite{Tashiro_2006,Oxyg_ISTLisbon} \\
R9 &
  $\mathrm{e + O_2(a) \rightarrow e + O(^3P) + O(^1D)}$ &
  $f(E/N)$ &
  1.27 &
  \cite{Phelps_1985_crossSections,Oxyg_ISTLisbon} \\
R10 &
  $\mathrm{e + O_2(b) \rightarrow e + 2O(^3P)}$ &
  $f(E/N)$ &
  0.84 &
  \cite{Tashiro_2006,Oxyg_ISTLisbon} \\
R11 &
  $\mathrm{e + O_2(b) \rightarrow e + O(^3P) + O(^1D)}$ &
  $f(E/N)$ &
  1.27 &
  \cite{Phelps_1985_crossSections,Oxyg_ISTLisbon} \\
R12 &
  $\mathrm{e + O_3 \rightarrow e + O(^3P) + O_2(X,0)}$ &
  $f(E/N)$ &
  4.9 &
  \cite{Gousset_1991,Oxyg_ISTLisbon} \\
\multicolumn{2}{l}{\textit{Electron-impact   ionization}} &
   &
   &
   \\
R13 &
  $\mathrm{e + O_2(X,0) \rightarrow 2e + O_2^+}$ &
  $f(E/N)$ &
  $-$ &
  \cite{Phelps_1985_crossSections,Oxyg_ISTLisbon} \\
R14 &
  $\mathrm{e + O_2(a) \rightarrow 2e + O_2^+}$ &
  $f(E/N)$ &
  $-$ &
  \cite{Gousset_1991,Oxyg_ISTLisbon} \\
R15 &
  $\mathrm{e + O(^3P) \rightarrow 2e + O^+}$ &
  $f(E/N)$ &
  $-$ &
  \cite{Laher_1990,Oxyg_ISTLisbon} \\
R16 &
  $\mathrm{e + O^- \rightarrow 2e + O(^3P)}$ &
  $f(E/N)$ &
  $-$ &
  \cite{Vejby-Christensen_1996,Oxyg_ISTLisbon} \\
\multicolumn{2}{l}{\textit{Electron-impact   dissociative ionization}} &
   &
   &
   \\
R17 &
  $\mathrm{e + O_2(X,0) \rightarrow 2e + O(^3P) + O^+}$ &
  $f(E/N)$ &
  $-$ &
  \cite{Gousset_1991,Oxyg_ISTLisbon} \\
R18 &
  e + O$_2$(a) $\rightarrow$ 2e + O($^3$P) + O$^+$ &
  $f(E/N)$ &
  $-$ &
  \cite{Gousset_1991,Oxyg_ISTLisbon} \\
\multicolumn{2}{l}{\textit{Electron-impact   dissociative attachment}} &
   &
   &
   \\
R19 &
  $\mathrm{e + O_2(X,0) \rightarrow O^- + O(^3P)}$ &
  $f(E/N)$ &
  1.27 &
  \cite{Phelps_1985_crossSections,Oxyg_ISTLisbon} \\
R20 &
  $\mathrm{e + O_2(a) \rightarrow O^- + O(^3P)}$ &
  $f(E/N)$ &
  1.27 &
  \cite{Burrow_1973,Oxyg_ISTLisbon} \\
\multicolumn{2}{l}{\textit{Electron-impact   dissociative recombination}} &
   &
   &
   \\
R21$^*$ &
  $\mathrm{e + O_2^+ \rightarrow 2O(^3P)}$ &
  $0.36\times2.2\times10^{-14} (T_e[\mathrm{eV}])^{-0.5}$ &
  6.98 &
  \cite{Gudmundsson_2004,Gudmundsson_2007,Petrignani_2005} \\
R22$^*$ &
  $\mathrm{e + O_2^+ \rightarrow O(^3P) + O(^1D)}$ &
  $0.64\times2.2\times10^{-14} (T_e[\mathrm{eV}])^{-0.5}$ &
  5.01 &
  \cite{Gudmundsson_2004,Gudmundsson_2007,Petrignani_2005} \\ \bottomrule
\end{tabular}
\end{table}

%% file: oxygenHeavySpeciesScheme.in
% \usepackage{booktabs}
% \usepackage{graphicx}
\begin{table}[]
\centering
\caption{List of heavy-species processes considered in this work for vibrationally-cold oxygen. The decay frequencies are in $\mathrm{s^{-1}}$ and the rate coefficients for two- and three-body reactions are in $\mathrm{m^3s^{-1}}$ and $\mathrm{m^6s^{-1}}$, respectively. New processes compared to \cite{Annusova_2018} are marked with **.  Processes with rate coefficients updated in this work, with respect to \cite{Annusova_2018},  are marked with *. Processes whose rate coefficients had a typo in \cite{Annusova_2018} are marked with \#.  $\epsilon_\mathrm{gh}$ is the energy released for gas heating and $f_\mathrm{wall}$ is the fraction of energy released in wall reactions that returns to the volume. $\mathrm{O_2(X)}$ includes all vibrational states of the ground-state of molecular oxygen, $\mathrm{O_2(X,0:41)}$.}
\label{tab:heavyReactions}
\resizebox{\textwidth}{!}{%
\scriptsize
\begin{tabular}{@{}lllll@{}}
\toprule
Reaction &
   &
  Rate coefficient   ($\mathrm{s^{-1}, m^3 s^{-1}, m^6 s^{-1}}$) &
  $\epsilon_\mathrm{gh}$ (eV) &
  Ref. \\ \midrule
\multicolumn{2}{l}{\textit{Neutral   species collisions}} &
   &
   &
   \\
R23$^\#$ &
  $\mathrm{O_2(a) + O(^3P) \rightarrow O_2(X,0) + O(^3P)}$ &
  $7\times10^{-22}$ &
  0.98 &
  \cite{Kossyi_1992,Guerra_1999} \\
R24$^{**}$ &
  $\mathrm{O_2(a) + O(^3P) + O_2(X) \rightarrow O_2(X,0) +   O(^3P) + O_2(X)}$ &
  $3\times10^{-44}$ &
  0.98 &
  \cite{Braginskiy_2005} \\
R25 &
  $\mathrm{O_2(a) + O_2(a) \rightarrow O_2(b) + O_2(X,0)}$ &
  $1.81\times10^{-24}\times(T_\mathrm{g}/300)^{3.8}\exp(700/T_\mathrm{g})$ &
  0.33 &
  \cite{Heidner_1981} \\
R26 &
  $\mathrm{O_2(a) + O_2(X) \rightarrow O_2(X,0) + O_2(X)}$ &
  $2.2\times10^{-24}(T_\mathrm{g}/300)^{0.8}$ &
  0.98 &
  \cite{Kossyi_1992} \\
R27 &
  $\mathrm{O_2(b) + O(^3P) \rightarrow O_2(a) + O(^3P)}$ &
  $4\times10^{-20}$ &
  0.65 &
  \cite{Ionin_2007} \\
R28$^*$ &
  $\mathrm{O_2(b) + O(^3P) \rightarrow O_2(X,0) + O(^3P)}$ &
  $4\times10^{-20} + 10^{-16}\exp(-3700/T_\mathrm{g})$ &
  1.63 &
  \cite{Booth_2022,Ionin_2007} \\
R29$^{**}$ &
  $\mathrm{O_2(b) + O(^3P) \rightarrow O_2(X,0) + O(^1D)}$ &
  $3.39\times10^{-17}\times(300/T_\mathrm{g})^{0.1}\exp(-4201/T_\mathrm{g})$ &
  -0.34 &
  \cite{Zinn_1982} \\
R30$^{**}$ &
  $\mathrm{O_2(Hz) + O(^3P) \rightarrow O(^3P) + O_2(X,0)}$ &
  $4.95\times10^{-18}$ &
  4.5 &
  \cite{Vasiljeva_2004} \\
R31$^{**}$ &
  $\mathrm{O_2(Hz) + O(^3P) \rightarrow O(^1D) + O_2(a)}$ &
  $2.7\times10^{-18}$ &
  1.55 &
  \cite{Vasiljeva_2004} \\
R32$^{**}$ &
  $\mathrm{O_2(Hz) + O(^3P) \rightarrow O(^1D) + O_2(b)}$ &
  $1.35\times10^{-18}$ &
  0.9 &
  \cite{Vasiljeva_2004} \\
R33$^{**}$ &
  $\mathrm{O_2(Hz) + O_2(X) \rightarrow O_2(X,0) + O_2(X)}$ &
  $0.23\times10^{-19}$ &
  4.5 &
  \cite{Vasiljeva_2004} \\
R34$^{**}$ &
  $\mathrm{O_2(Hz) + O_2(X) \rightarrow O_2(a) + O_2(X)}$ &
  $1.86\times10^{-19}$ &
  3.52 &
  \cite{Vasiljeva_2004} \\
R35$^{**}$ &
  $\mathrm{O_2(Hz) + O_2(X) \rightarrow O_2(b) + O_2(X)}$ &
  $0.21\times10^{-19}$ &
  2.87 &
  \cite{Vasiljeva_2004} \\
R36 &
  $\mathrm{O(^1D) + O(^3P) \rightarrow 2O(^3P)}$ &
  $8\times10^{-18}$ &
  1.97 &
  \cite{Eliasson_1986} \\
R37 &
  $\mathrm{2O(^3P) + O_2(X) \rightarrow O_2(X,0) + O_2(X)}$ &
  $0.5\times3.81\times10^{-42}T_\mathrm{g}^{-1}\exp(-170/T_\mathrm{g})$ &
  5.16 &
  \cite{Eliasson_1986,Feoktistov_1995} \\
R38 &
  $\mathrm{2O(^3P) + O_2(X) \rightarrow O_2(a) + O_2(X)}$ &
  $0.33\times3.81\times10^{-42}T_\mathrm{g}^{-1}\exp(-170/T_\mathrm{g})$ &
  4.18 &
  \cite{Eliasson_1986,Feoktistov_1995} \\
R39 &
  $\mathrm{2O(^3P) + O_2(X) \rightarrow O_2(b) + O_2(X)}$ &
  $0.17\times3.81\times10^{-42}T_\mathrm{g}^{-1}\exp(-170/T_\mathrm{g})$ &
  3.53 &
  \cite{Eliasson_1986,Feoktistov_1995} \\
R40$^*$ &
  $\mathrm{3O(^3P) \rightarrow O_2(X,0) + O(^3P)}$ &
  $2.5\times10^{-43}T_\mathrm{g}^{-0.63}$ &
  5.16 &
  \cite{Gordiets_1995} \\
R41 &
  $\mathrm{O(^1D) + O_2(X) \rightarrow O(^3P) + O_2(X)}$ &
  $7\times10^{-18}\exp(67/T_\mathrm{g})$ &
  1.97 &
  \cite{Eliasson_1986} \\
R42$^*$ &
  $\mathrm{O(^1D) + O_2(X,0) \rightarrow O(^3P) + O_2(a)}$ &
  $8\times10^{-19}\exp(67/T_\mathrm{g})$ &
  0.99 &
  \cite{Niles_1974,Atkinson_2004} \\
R43 &
  $\mathrm{O(^1D) + O_2(X,0) \rightarrow O(^3P) + O_2(b)}$ &
  $2.56\times10^{-17}\exp(67/T_\mathrm{g})$ &
  0.34 &
  \cite{Eliasson_1986} \\
R44 &
  $\mathrm{O_2(b) + O_3 \rightarrow 2O_2(X,0) + O(^3P)}$ &
  $1.5\times10^{-17}$ &
  0.53 &
  \cite{Eliasson_1986} \\
R45 &
  $\mathrm{O(^1D) + O_3 \rightarrow 2O_2(X,0)}$ &
  $1.2\times10^{-16}$ &
  6.03 &
  \cite{Eliasson_1986} \\
R46 &
  $\mathrm{O(^1D) + O_3 \rightarrow O_2(X,0) + 2O(^3P)}$ &
  $1.2\times10^{-16}$ &
  0.87 &
  \cite{Eliasson_1986} \\
R47 &
  $\mathrm{O_2(a) + O_3 \rightarrow 2O_2(X,0) + O(^3P)}$ &
  $5.2\times10^{-17}\exp(-2840/T_\mathrm{g})$ &
  -0.12 &
  \cite{Eliasson_1986} \\
R48 &
  $\mathrm{O(^3P) + O_3 \rightarrow 2O_2(X,0)}$ &
  $0.5\times1.8\times10^{-17}\exp(-2300/T_\mathrm{g})$ &
  4.06 &
  \cite{Eliasson_1986,Feoktistov_1995} \\
R49$^\#$ &
  $\mathrm{O(^3P) + O_3 \rightarrow O_2(a) + O_2(X,0)}$ &
  $0.33\times1.8\times10^{-17}\exp(-2300/T_\mathrm{g})$ &
  3.08 &
  \cite{Eliasson_1986,Feoktistov_1995} \\
R50$^\#$ &
  $\mathrm{O(^3P) + O_3 \rightarrow O_2(b) + O_2(X,0)}$ &
  $0.17\times1.8\times10^{-17}\exp(-2300/T_\mathrm{g})$ &
  2.43 &
  \cite{Eliasson_1986,Feoktistov_1995} \\
R51 &
  $\mathrm{O_3^* + O(^3P) \rightarrow O_3 + O(^3P)}$ &
  $2\times10^{-19}$ &
  0.09 &
  \cite{Lopaev_2011} \\
R52 &
  $\mathrm{O_3^* + O_2(X) \rightarrow O_3 + O_2(X)}$ &
  $3\times10^{-21}$ &
  0.09 &
  \cite{Lopaev_2011} \\
R53 &
  $\mathrm{O_2(a) + O_3^* \rightarrow 2O_2(X,0) + O(^3P)}$ &
  $5\times5.2\times10^{-17}\exp(-(2840-1553)/T_\mathrm{g})$ &
  -0.03 &
  \cite{Marinov_2013} \\
R54 &
  $\mathrm{O(^3P) + O_3^* \rightarrow 2O_2(X,0)}$ &
  $8\times10^{-18}\exp(-(2060-1553)/T_\mathrm{g})$ &
  4.15 &
  \cite{Lopaev_2011} \\
R55 &
  $\mathrm{2O(^3P) + O_2(X,0) \rightarrow O_3 + O(^3P)}$ &
  $2.1\times10^{-46}\exp(345/T_\mathrm{g})$ &
  1.1 &
  \cite{Eliasson_1986} \\
R56 &
  $\mathrm{O(^3P) + O_2(X,0) + O_2(X) \rightarrow O_3 + O_2(X)}$ &
  $0.33\times6.4\times10^{-47}\exp(663/T_\mathrm{g})$ &
  1.1 &
  \cite{Eliasson_1986,Lopaev_2011} \\
R57$^\#$ &
  $\mathrm{O(^3P) + O_2(X,0) + O_2(X) \rightarrow O_3^* +   O_2(X)}$ &
  $0.67\times6.4\times10^{-47}\exp(663/T_\mathrm{g})$ &
  1.01 &
  \cite{Eliasson_1986,Lopaev_2011} \\
R58 &
  $\mathrm{O(^3P) + O_2(X,0) + O_3 \rightarrow  2O_3}$ &
  $1.66\times10^{-46}\exp(T_\mathrm{g}/300)$ &
  1.1 &
  \cite{Belmonte_1997} \\
\multicolumn{2}{l}{\textit{Ion   collisions}} &
   &
   &
   \\
R59 &
  $\mathrm{O^- + O_2(X,0) \rightarrow O_3 + e}$ &
  $10^{-18}$ &
  $-$ &
  \cite{Fridman_2008} \\
R60 &
  $\mathrm{O^- + O_2(a) \rightarrow O_3 + e}$ &
  $0.75\times1.9\times10^{-16}$ &
  $-$ &
  \cite{Eliasson_1986} \\
R61 &
  $\mathrm{O^- + O(^3P) \rightarrow O_2(X,0) + e}$ &
  $1.3\times10^{-15}$ &
  $-$ &
  \cite{Fridman_2008} \\
R62 &
  $\mathrm{O^- + O_2(b) \rightarrow O(^3P) + O_2(X,0) + e}$ &
  $6.9\times10^{-16}$ &
  $-$ &
  \cite{Eliasson_1986} \\
R63 &
  $\mathrm{O^+ + O^- \rightarrow 2O(^3P)}$ &
  $2.8\times10^{-13}$ &
  12.15 &
  \cite{Olson_1970} \\
R64 &
  $\mathrm{O^+  + O_3   \rightarrow O_2^+ + O_2(X,0)}$ &
  $10^{-16}$ &
  5.6 &
  \cite{Kossyi_1992} \\
R65 &
  $\mathrm{O^+ + O_2(X,0) \rightarrow O_2^+ + O(^3P)}$ &
  $2\times10^{-17}(300/T_\mathrm{g})^{0.5}$ &
  1.54 &
  \cite{Eliasson_1986} \\
R66 &
  $\mathrm{O^+ + O_2(a) \rightarrow O_2^+ + O(^3P)}$ &
  $2\times10^{-17}(300/T_\mathrm{g})^{0.5}$ &
  2.52 &
  \cite{Eliasson_1986,Annusova_2018} \\
R67 &
  $\mathrm{O_2^+ + O^- \rightarrow O_2(X,0) + O(^3P)}$ &
  $9.6\times10^{-14}(300/T_\mathrm{g})^{0.5}$ &
  10.61 &
  \cite{Eliasson_1986} \\
\multicolumn{2}{l}{\textit{Ion   transport}} &
   &
   &
   \\
R68 &
  $\mathrm{O_2^+ + wall \rightarrow O_2(X,0) + wall}$ &
  Classic. ambip. diff. with neg. ions &
  $f_\mathrm{wall}\times12.14$ &
  \cite{Guerra_1999} \\
R69 &
  $\mathrm{O^+ + wall \rightarrow O(^3P) + wall}$ &
  Classic. ambip. diff. with neg. ions &
  $f_\mathrm{wall}\times13.68$ &
  \cite{Guerra_1999} \\
\multicolumn{2}{l}{\textit{Neutral   transport}} &
   &
   &
   \\
R70$^*$ &
  $\mathrm{O_2(a) + wall \rightarrow O_2(X,0) + wall}$ &
  $\gamma_\mathrm{O_2(a)} = 2.2\times10^{-4}$ &
  $f_\mathrm{wall}\times0.98$ &
  \cite{Booth_2020} \\
R71$^*$ &
  $\mathrm{O_2(b) + wall \rightarrow O_2(X,0) + wall}$ &
  $\gamma_\mathrm{O_2(b)} = 0.135$ &
  $f_\mathrm{wall}\times1.63$ &
  \cite{Booth_2022} \\
R72$^{**}$ &
  $\mathrm{O_2(Hz) + wall \rightarrow O_2(X,0) + wall}$ &
  $\gamma_\mathrm{O_2(Hz)} = 1$ &
  $f_\mathrm{wall}\times4.5$ &
  \cite{Vasiljeva_2004} \\
R73 &
  $\mathrm{O(^3P) + wall \rightarrow \frac{1}{2}O_2(X,0) +   wall}$ &
  $\gamma_\mathrm{O(^3P)} = f(p,I,T_w)$ &
  $f_\mathrm{wall}\times2.58$ &
  \cite{Booth_2020} \\
R74 &
  $\mathrm{O(^1D) + wall \rightarrow O(^3P) + wall}$ &
  $\gamma_\mathrm{O(^1D)} = 1$ &
  $f_\mathrm{wall}\times1.97$ &
  \cite{Gordiets_1995} \\
R75 &
  $\mathrm{O_3^* + wall \rightarrow O_3 + wall}$ &
  $\gamma_\mathrm{O_3^*} = 0.1$ &
  $f_\mathrm{wall}\times0.09$ &
  \cite{Lopaev_2011,Marinov_2013} \\ \bottomrule
\end{tabular}%
}
\pagebreak
\end{table}

%% file: oxygenVibrationsScheme.in
\begin{table}[]
\centering
\scriptsize
\caption{List of e-V, V-T and V-V processes considered in this work to describe the vibrational kinetics of $\mathrm{O_2(X,v)}$. $\epsilon_\mathrm{gh}$ is the energy released for gas heating and $f_\mathrm{wall}$ is the fraction of energy released in wall reactions that returns to the volume. ``$\sigma_{76}$ scaled from $\sigma_1$'' means that the electron-scattering cross-section describing R76 ($\sigma_{76}$) is scaled from the cross-section describing R1 ($\sigma_1$). The same for the other cases. The scaling procedures are referred in the text of section~\ref{sec:vibKin}.}
\label{tab:vibKinetics}
\begin{tabular}{@{}lllll@{}}
\toprule
Reaction &
   &
  Rate coefficient   ($\mathrm{s^{-1}, m^3 s^{-1}}$) &
  $\epsilon_\mathrm{gh}$ (eV) &
  Ref. \\ \midrule
\multicolumn{2}{l}{\textit{Electron-impact   excitation}} &
   &
   &
   \\
R76 &
  $\mathrm{e + O_2(X,1:6) \rightarrow e + O_2(a)}$ &
  $f(E/N)$, $\sigma_{76}$ scaled from $\sigma_1$ &
  $-$ &
  \cite{Annusova_2018,Oxyg_ISTLisbon} \\
R77 &
  $\mathrm{e + O_2(X,1:8) \rightarrow e + O_2(b)}$ &
  $f(E/N)$, $\sigma_{77}$ scaled from $\sigma_2$ &
  $-$ &
  \cite{Annusova_2018,Oxyg_ISTLisbon} \\
R78 &
  $\mathrm{e + O_2(X,v) \leftrightarrow    e + O_2(X,w), \Delta v = v-w = 1:30}$ &
  $f(E/N)$ &
  $-$ &
  \cite{Laporta_2013,Oxyg_ISTLisbon} \\
\multicolumn{2}{l}{\textit{Electron-impact   dissociation}} &
   &
   &
   \\
R79 &
  $\mathrm{e + O_2(X,1:41) \rightarrow e + O(^3P) + O(^3P)}$ &
  $f(E/N)$, $\sigma_{79}$ scaled from $\sigma_6$ &
  0.84 &
  \cite{Laporta_2015,Annusova_2018} \\
R80 &
  $\mathrm{e + O_2(X,1:41) \rightarrow e + O(^3P) + O(^1D)}$ &
  $f(E/N)$, $\sigma_{80} = \sigma_7$ &
  1.27 &
  \cite{Phelps_1985_crossSections,Annusova_2018} \\
\multicolumn{2}{l}{\textit{Electron-impact   ionization}} &
   &
   &
   \\
R81 &
  $\mathrm{e + O_2(X,1:41) \rightarrow e + e + O_2^+}$ &
  $f(E/N)$, $\sigma_{81} = \sigma_{13}$ &
  $-$ &
  \cite{Phelps_1985_crossSections,Annusova_2018} \\
\multicolumn{2}{l}{\textit{Electron-impact   dissociative attachment}} &
   &
   &
   \\
R82 &
  $\mathrm{e + O_2(X,1:41) \rightarrow e +  O^-    + O(^3P)}$ &
  $f(E/N)$, $\sigma_{82}$ scaled from $\sigma_{19}$ &
  1.27 &
  \cite{Laporta_2015,Annusova_2018} \\
\multicolumn{2}{l}{\textit{Vibration-to-translation   relaxation}} &
   &
   &
   \\
R83 &
  $\mathrm{O_2(X,v) + O \leftrightarrow O_2(X,w) + O}$ &
  $f(\mathrm{v,w},T_\mathrm{g})$ &
  $E_\mathrm{v}-E_\mathrm{w}$ &
  \cite{Esposito_2008} \\
R84 &
  $\mathrm{O_2(X,v) + O_2(X) \leftrightarrow O_2(X,v-1) +   O_2(X)}$ &
  $f(\mathrm{v},T_\mathrm{g})$ &
  $E_\mathrm{v}-E_\mathrm{v-1}$ &
  \cite{LinodaSilva_2012,Annusova_2018} \\
\multicolumn{2}{l}{\textit{Vibration-to-vibration   transfer}} &
   &
   &
   \\
R85 &
  $\mathrm{O_2(X,v) + O_2(X,w-1) \leftrightarrow O_2(X,v-1) +   O_2(X,w)}$ &
  $f(\mathrm{v,w},T_\mathrm{g})$ &
  $E_\mathrm{v}+E_\mathrm{w-1}-E_\mathrm{v-1}-E_\mathrm{w}$ &
  \cite{LinodaSilva_2012,Annusova_2018} \\
\multicolumn{2}{l}{\textit{Wall   de-excitation}} &
   &
   &
   \\
R86 &
  $\mathrm{O_2(X,v) + wall \rightarrow O_2(X,v-1) + wall}$ &
  $\gamma_\mathrm{v} = 1.1\times10^{-3}$ &
  $f_\mathrm{wall}\times(E_\mathrm{v}-E_\mathrm{v-1})$ &
  \cite{Marinov_2012,Annusova_2018} \\ \bottomrule
\end{tabular}
\end{table}